\documentclass{emulateapj}
\pdfoutput=1

\shorttitle{SNe Ia from Merging WDs I) Prompt Detonations}
\shortauthors{Moll et al.}

\usepackage{amsmath}
\usepackage{natbib}
\usepackage{multirow}
\usepackage{xspace}
\newcommand{\mnine}{\mbox{0.96\,+\,0.81}\xspace}
\newcommand{\mone}{\mbox{1.06\,+\,1.06}\xspace}
\newcommand{\mtwo}{\mbox{1.20\,+\,1.06}\xspace}
\newcommand{\abs}[1]{\ensuremath{\left|#1\right|}}

\newcommand{\unit}[1]{\ensuremath{\;\mathrm{#1}}}
\newcommand{\gcc}{\ensuremath{\unit{g\;cm^{-3}}}}
\newcommand{\cms}{\ensuremath{\unit{cm\;s^{-1}}}}
\newcommand{\kms}{\ensuremath{\unit{km\;s^{-1}}}}
\newcommand{\ergss}{\ensuremath{\unit{ergs\;s^{-1}}}}
\newcommand{\degree}{\ensuremath{^\circ}}
\newcommand{\simm}{\ensuremath{\mathord{\sim}}}
\newcommand{\de}{\ensuremath{\mathrm{d}}}
\newcommand{\el}[2]{$^{#2}\text{#1}$}
\newcommand{\imagetop}[1]{\vtop{\null\hbox{#1}}}
\newcommand{\Nifs}{\el{Ni}{56}\xspace}
\newcommand{\Msun}{\ensuremath{M_\odot}}
\newcommand{\Lbol}{\ensuremath{L_\text{bol}}}
\renewcommand{\vec}[1]{\boldsymbol{#1}}
\newcommand{\myscale}{1}

\begin{document}

\title{Type Ia Supernovae from Merging White Dwarfs I) Prompt Detonations}

\author{R.~Moll\altaffilmark{1,4}, C.~Raskin\altaffilmark{2}, D.~Kasen\altaffilmark{2,3} and S.~E.~Woosley\altaffilmark{1}}

\altaffiltext{1}{Department of Astronomy and Astrophysics, University of California, Santa Cruz, CA 95064, USA}
\altaffiltext{2}{Nuclear Science Division, Lawrence Berkeley National Laboratory, Berkeley, CA 94720, USA}
\altaffiltext{3}{Department of Physics and Astronomy, University of California, Berkeley, CA 94720, USA}
\altaffiltext{4}{Max-Planck-Institut f\"{u}r Astrophysik, Karl-Schwarzschild-Str.~1, 85748~Garching, Germany}

\date{\today}

\begin{abstract}
Merging white dwarfs are a possible progenitor of Type Ia supernovae
(SNe~Ia).  While it is not entirely clear if and when an explosion is
triggered in such systems, numerical models suggest that a detonation
might be initiated before the stars have coalesced to form a single
compact object.  Here we study such ``peri-merger'' detonations by
means of numerical simulations, modeling the disruption and
nucleosynthesis of the stars until the ejecta reach the coasting
phase. Synthetic light curves and spectra are generated for comparison
with observations.  Three models are considered with primary masses
$0.96\,\Msun$, $1.06\,\Msun$, and $1.20\,\Msun$.  Of these, the
$0.96\,\Msun$ dwarf merging with an $0.81\,\Msun$ companion, with a
\Nifs yield of $0.58\,\Msun$, is the most promising candidate for
reproducing common SNe~Ia.  The more massive mergers produce unusually
luminous SNe~Ia with peak luminosities approaching those attributed to
``super-Chandrasekhar'' mass SNe~Ia.  While the synthetic light curves
and spectra of some of the models resemble observed SNe~Ia, the
significant asymmetry of the ejecta leads to large orientation
effects.  The peak bolometric luminosity varies by more than a factor
of 2 with the viewing angle, and the velocities of the spectral
absorption features are lower when observed from angles where the
light curve is brightest.  The largest orientation effects are seen in
the ultraviolet, where the flux varies by more than an order of
magnitude.  Despite the large variation with viewing angle, the set of
three models roughly obeys a width-luminosity relation, with the
brighter light curves declining more slowly in the B-band.  Spectral
features due to unburned carbon from the secondary star are also seen
in some cases.
\end{abstract}


\section{Introduction}
\label{sec:intro}

It is generally believed that the progenitors of type Ia supernovae
(SNe~Ia) are the explosion of carbon-oxygen white dwarfs (WDs) powered
by thermonuclear fusion.  There is currently no consensus, however,
regarding the events leading to the explosion.  Models of SNe~Ia
progenitor systems are generally classified as either
``single-degenerate'' or ``double-degenerate'', depending on whether
the progenitor system consists of one WD and a non-degenerate
companion, or two WDs. Single-degenerate scenarios are further divided
into Chandrasekhar mass models \citep[for a review,
  see][]{2000Hillebrandt} and sub-Chandrasekhar mass models
\citep[e.g.][]{2011Woosley}, depending on the mass of the exploding
star.  Detonations of sub-Chandrasekhar mass WDs may be triggered by
the explosion of a helium shell, which ignites more readily than the
carbon-oxygen mixture out of which the core WD is assumed to be
composed \citep{2007Fink,2010Fink,2012Sim,2013Shen,2013Moll}.

In the double-degenerate scenario, two WDs merge and explode, an idea
suggested 30 years ago as a possible progenitor for SNe~Ia
\citep{1984Iben,1984Webbink}.  This scenario can naturally explain the lack
of hydrogen spectral lines in SNe~Ia, and is supported by studies of stellar
population synthesis \citep{2011Ruiter,2013Ruiter}.  The merger starts when
the lighter of the two stars overflows its Roche lobe.  The ensuing mass
transfer is often unstable \citep{2004Marsh}, leading to the rapid disruption
of the less massive star \citep[e.g.,][]{1990Benz,2007Yoon}.

Recent simulations have shown that detonations leading to SNe~Ia could
be triggered relatively early in the merging process.
\citet{2011Pakmor} explored the coalescence of
$\mathord{\approx}0.90\,\Msun$ WDs with secondaries between
$0.70\,\Msun$ and $0.89\,\Msun$. For secondary masses down to
$0.76\,\Msun$, they found conditions favorable for a detonation during
the early phases of the merger. In the case of two $0.89\,\Msun$ WDs,
they followed the merger, detonation and the subsequent explosion,
finding that the model synthesized a relatively small amount of \Nifs
and resembled the class of sub-luminous, 1991bg-like SNe~Ia
\citep{2010Pakmor}.  Mergers that included a higher mass WD, $M \simeq
1.1\,\Msun$, produced more \Nifs and might yield  a more normal SN~Ia
\citep{2012Pakmor,2012bPakmor}.  Models of a $(1.1+0.9)\Msun$ merger
provided a reasonable match to the light curves and spectra of SN
2011fe \citep{2012Roepke}.

The possibility of such ``peri-merger detonations'' (i.e., detonations
set off during the merging process) is disputed in the literature.
\citet{2004Guerrero} conclude that the formation of an SN~Ia during
the merger is unlikely.  The parameter study of \citet{2012Dan} also
argues against such an outcome in the case of pure carbon-oxygen WDs.
However, the resolution in these studies was low compared to other
recent merger calculations \citep{2012Raskin,2013Zhu}, and this has
likely played a role in the disparate conclusions.
If the merger does not lead to an early thermonuclear runaway, it will
ultimately make a single compact object surrounded by a disk
\citep[e.g.,][]{2009Loren}. We shall consider the possibility of explosions in
this post-merger state in a separate paper \citep{Raskin_2013}.

In this work, we study the
possibility and consequences of peri-merger detonations. 
\S2 describes the merger setups and our detonation
conditions. \S3 presents the resulting yields and ejecta morphology.
\S4 discusses radiative transport calculations of the synthetic
light curves and spectra carried out for azimuthally averaged models.
\S4 summarizes our
findings and discusses avenues for further study.

\section{Methods}
\label{sec:methods}

\subsection{Merging and Initial Models}
\label{sec:inimodels}

\renewcommand{\myscale}{.47}
\begin{figure*}[t]
\centering
\begin{tabular}{lll}
a) \hspace{-12pt} \imagetop{ \includegraphics[scale=\myscale]{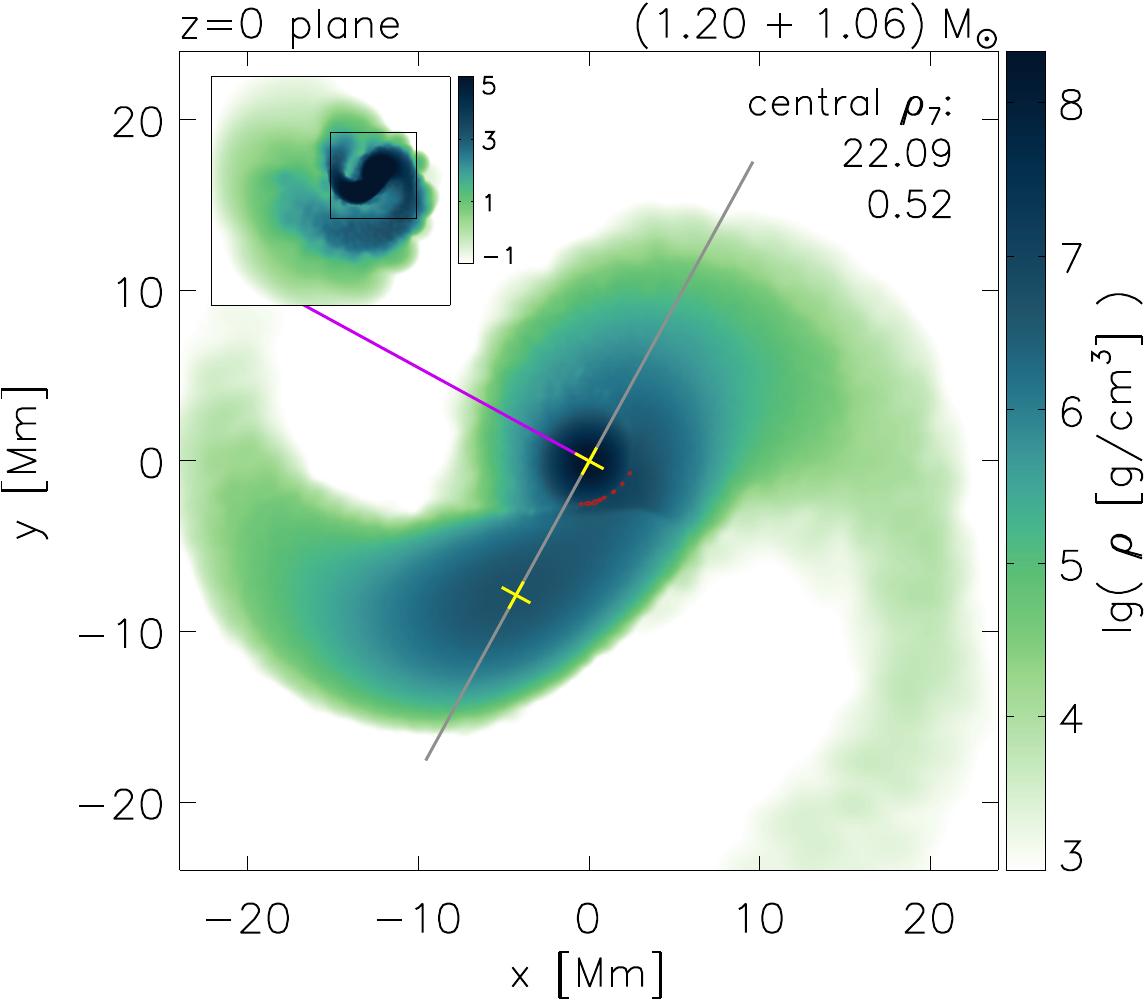} } &
b) \hspace{-12pt} \imagetop{ \includegraphics[scale=\myscale]{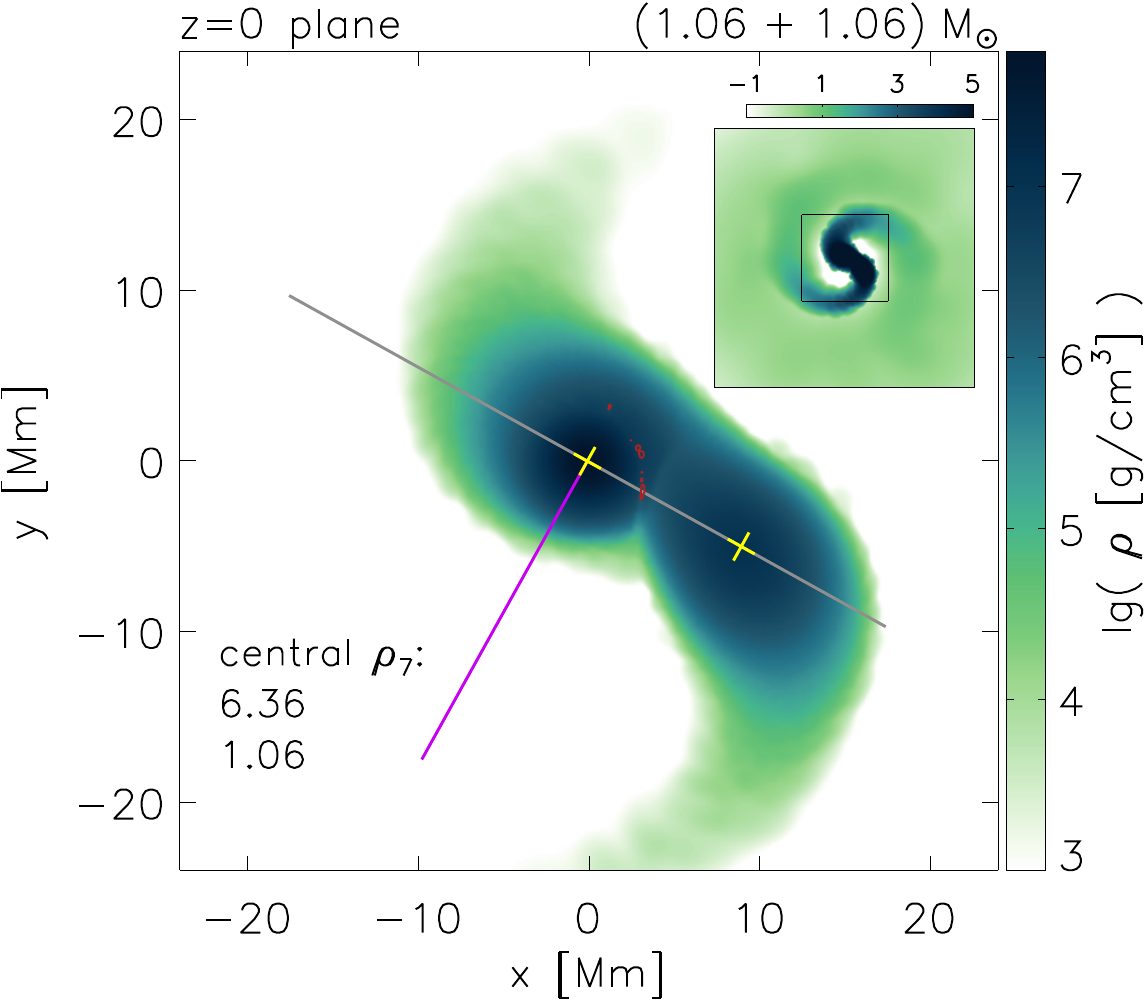} } &
c) \hspace{-12pt} \imagetop{ \includegraphics[scale=\myscale]{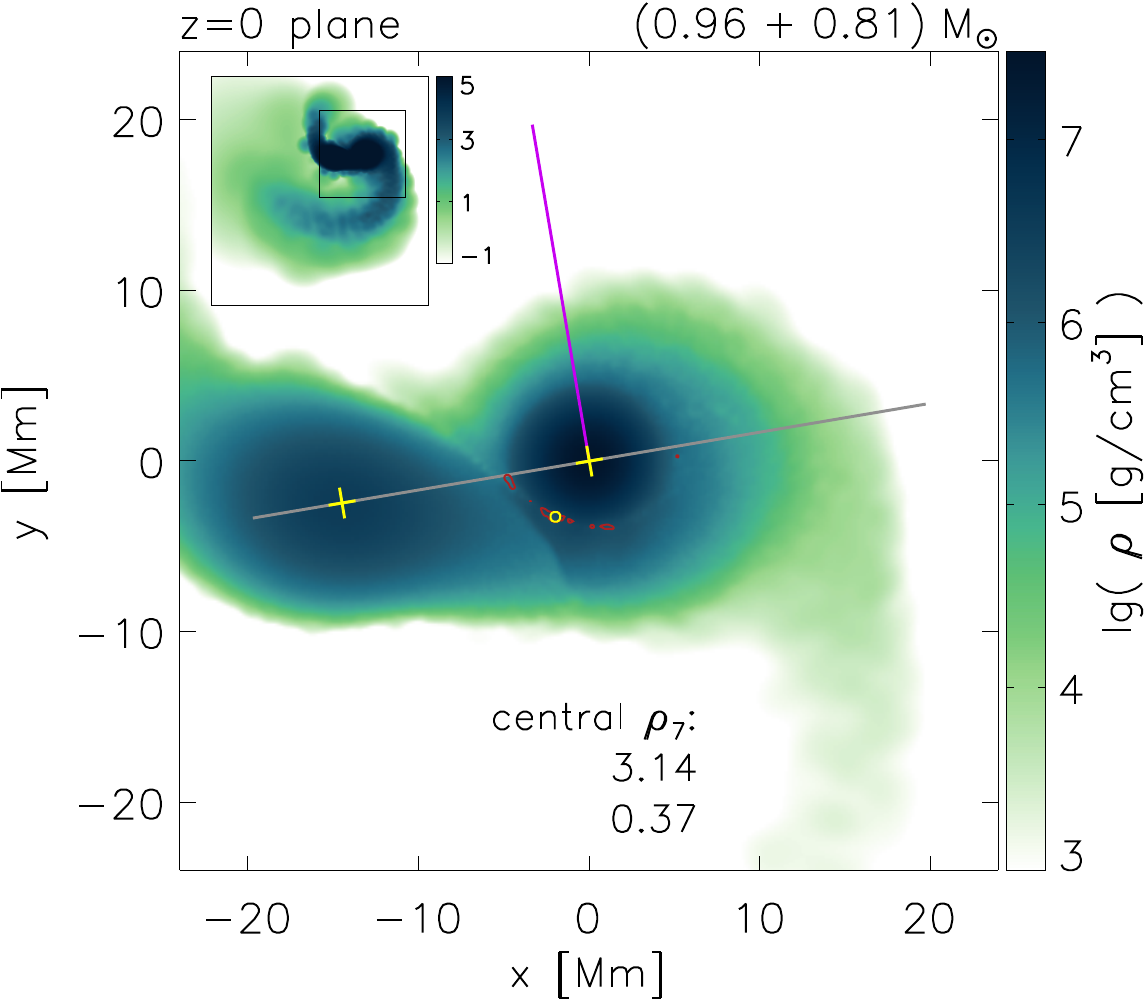} }
\end{tabular}
\caption{Density in the orbital plane at the beginning of the detonation
simulations.  Yellow crosses mark the density maxima of the two stars, which
for clarity are also written inside each plot (in units of
$10^7\gcc$). The denser primary is centered at the coordinate origin. Red
contours indicate the hottest regions in the plane (the respective levels are
$T_9=2.0$, $1.2$, $0.8$ for panels a, b, and c).  The small yellow circle in
panel (c) represents the perimeter of the detonator (contours at $T_9=2.0$)
which is needed to get a detonation going in this model.  The insets in each
panel show a larger region, with the small black squares indicating the
boundaries of the respective main plot. The gray and magenta lines indicate
axes in the orbital plane that are used in the description of the results.}
\label{fig:initrho}
\end{figure*}

\begin{deluxetable}{c@{}r|rrr}
\tablecaption{Masses [$\Msun$] at different densities
at the beginning of the detonation simulations}
\tablehead{ \multicolumn{2}{r}{$\rho\,[\mathrm{g\;cm^{-3}}]$ $\downarrow$} &
            \colhead{\mtwo} & \colhead{\mone} & \colhead{\mnine} }
\startdata
\multirow{5}{*}{\rotatebox{90}{primary}}
& $> 10^5$          & $\simm1.406$ & $\simm1.222$ & $\simm0.993$ \\
& $> 2\times 10^6$  & $\simm1.280$ & $\simm1.101$ & $0.853$ \\
& $> 5\times 10^6$  & $1.177$ & $0.922$ & $0.666$ \\
& $> 10^7$          & $1.069$ & $0.738$ & $0.430$ \\
& $> 2\times 10^7$  & $0.935$ & $0.475$ & $0.133$ \\\hline
\multirow{5}{*}{\rotatebox{90}{secondary}}
& $> 10^5$          & $\simm0.638$ & $\simm0.792$ & $\simm0.521$ \\
& $> 2\times 10^6$  & $\simm0.339$ & $\simm0.579$ & $0.160$ \\
& $> 5\times 10^6$  & $0.002$ & $0.257$ & $0.000$ \\
& $> 10^7$          & $0.000$ & $0.006$ & $0.000$
\enddata
\label{tab:initmasses}
\end{deluxetable}

To simulate the merger phase of the calculation, we use the smoothed
particle hydrodynamics code SNSPH \citep{2006Fryer}. Following
procedures laid out in \citet{2012Raskin} for generating initial
conditions, we ensure accurate mass transfer rates consistent with the
results of \citet{2012Dan}. When our merger simulations approach the
detonation conditions of \citet{2009Seitenzahl}, we interpolate an
earlier SNSPH snapshot onto a uniform Cartesian grid at three
different resolutions, corresponding to different levels of
refinement.  We then evolve the system with the Eulerian grid code
CASTRO (see the next section).  Figure~\ref{fig:initrho} presents an
overview of the snapshots that were followed up with CASTRO.  The
hottest regions are located near the orbital plane ($\abs{z} \lesssim
\pm 1\unit{Mm}$) where the accretion stream from the secondary star
impacts the surface of the primary, see the red contours in the plots.
Shortly after the simulations are restarted in CASTRO, detonations
spontaneously commence at one or several points in these regions.
Looking down the $z$-axis as in Figure~\ref{fig:initrho}, the stars
are rotating around a common center (close to the primary) in
counter-clockwise direction.  We refer to this general direction of
rotation in the description of the results when stating that something
is ``ahead'' or ``behind'' one of the stars.

Due to inconsistencies in the way SNSPH and CASTRO handle shock physics, the
precise times and locations where detonation conditions are reached are not
identical after the code-swap.  However, since an appreciable amount of
material is reaching detonation conditions at roughly the same time in the
SNSPH simulation---as is demonstrated by the red contours in the mapped CASTRO
snapshots in Figure~\ref{fig:initrho}---this slight variation should not affect
the outcome.  With the exception of model \mnine, the largest disparity in
times between detonation conditions being reached in the two codes is
$\simm5\unit{s}$.

One of the models, \mnine, does not ignite in CASTRO. After mapping the
SPH data onto the Cartesian grid, the hottest spot in the orbital plane has a
temperature of $1.26\times10^9\unit{K}$, and a density of $1.32\times10^6\gcc$
at the highest level of refinement, corresponding to $3.5\times10^{-8}\,\Msun$
in that zone.  The hottest point overall is $1069\unit{km}$ away from the
orbital plane, and has a temperature of $1.51\times10^9\unit{K}$, but the
density there is only $0.88\times10^6\gcc$.  These spots do not induce a
detonation in CASTRO, even with a full 19-isotope network (running with a full
network avoids the lower temperature limit of our tables), and no higher
temperatures develop in the course of the following 10 seconds in the orbital
plane.  As argued above, we expect the switch from SNSPH to CASTRO, and the
interpolation of the data onto a grid, to have an unavoidable impact on the
thermal evolution.  In light of this, we rate the possibility for ignition of
this model as uncertain.  To force a detonation, we manually put a hot spot
with a central temperature of $3\times10^9\unit{K}$ and a radius of
$300\unit{km}$ at the location of the hottest point in the orbital plane, see
the small yellow circle in Figure~\ref{fig:initrho}c.

The central densities indicate varying evolutionary states (degree of
deformation and mass loss) of the secondary stars at the time of the
detonation.  For example, the maximum density of the secondary in model \mtwo
is only about half as high as in the model \mone, even though they start out
with the same masses.  Table~\ref{tab:initmasses} lists the masses of gas above
a given density cutoff at the beginning of the detonation simulations. The
values represent the masses of contiguous regions about the centers of the
primary and secondary stars.  For low densities, the contiguous regions are
identical, and we split the total mass into parts associated with the primary
and the secondary by means of a cutting plane, the location of which is
determined by eye from the density in the orbital plane.  These split values
are preceded by a ``$\sim$'' in the tables. The numbers show that the density
of the material in the secondary is relatively low, and that a significant mass
transfer has taken place before the detonation.

Even though model \mone starts with two WDs of identical mass, a slight
geometric asymmetry is sufficient to unbind one of the stars, as was described
in \citet{2012Raskin}. Since perfectly identical stars in binaries are probably
exceedingly rare, we expect this outcome to be more likely in
nature.

\subsection{Detonation and Explosion}
\label{sec:hydro}

\begin{deluxetable}{lrrrr}
\tablecaption{\el{Ni}{56} yields [$\Msun$] for bare carbon-oxygen WDs with different codes and networks}
\tablehead{ & \colhead{$1.1\,\Msun$} & \colhead{$1.0\,\Msun$} & \colhead{$0.9\,\Msun$} & \colhead{$0.8\,\Msun$}}
\startdata
KEPLER 19-iso\tablenotemark{a}  & 0.791 & 0.488 & 0.166 & 0.0167 \\
CASTRO 19-iso                   & 0.769 & 0.473 & 0.160 & 0.0118 \\
KEPLER 199-iso\tablenotemark{b} & 0.825 & 0.566 & 0.287 & 0.0551 \\
CASTRO table\tablenotemark{c}   & 0.748 & 0.517 & 0.271 & 0.0668
\enddata
\tablenotetext{a}{using a 19-isotope network}
\tablenotetext{b}{using a 199-isotope network directly}
\tablenotetext{c}{using a table based on the 199-isotope network}
\label{tab:tabletest}
\end{deluxetable}

For our 3D detonation study, we use the Eulerian hydrodynamics code CASTRO
\citep{2010Almgren,2011Zhang} to solve the equations for compressible fluid
dynamics in combination with a nuclear reaction network. The equation of state
is based on the Helmholtz free energy \citep{1999Timmes,2000Timmes}, and the
gravitational potential is calculated with a Poisson solver. We use lookup
tables based on a 199-isotope network to compute the nuclear energy generation
and element synthesis, with the yields packed into 19 isotopes. In dense ($\rho
> 10^6\gcc$), carbon-rich ($X_\mathrm{C} > 0.1$) zones with temperatures
greater than $1.5\times10^9\unit{K}$, the composition changes and energy
release from carbon burning are determined from the table.  Once the
temperature has risen above $2.1\times10^9\unit{K}$ in a zone, another table is
used to determine the final, freeze-out composition. To prevent a spurious
propagation of the detonation through mixing with hot ash, this last step is
only allowed if burning conditions in a particular zone are met for at least
half a sound crossing time.  The accuracy of the tables in CASTRO was confirmed
on a set of bare carbon-oxygen WDs by comparison with the KEPLER code
\citep{2011Woosley}, see Table~\ref{tab:tabletest}.  The stars in these 1D
tests were resolved by approximately 160 radial zones in CASTRO, which is
roughly similar to the 3D resolution used for the merger detonations.  We find
reasonable agreement between 199-isotope network simulations in KEPLER and
simulations with the tables in CASTRO. The small network is much less efficient
in producing \el{Ni}{56} in low-mass WDs.

The detonation simulations are started with 3 levels of static mesh refinement.
Each level consists of a cube centered on the point of highest density. The
domain sizes are $(12\times10^8\unit{cm})^3$, $(48\times10^8\unit{cm})^3$ and
$(192\times10^8\unit{cm})^3$.  The grid is resolved with
$320\times320\times320$ zones at each level.  The innermost level is dropped
when the detonation reaches the refinement boundary. Once the ejecta gets near
outer boundary, we map the data into a new domain twice as large, dropping the
innermost level of refinement.  This step is repeated until the ejecta gets
near the boundary of a domain of $(3.07\times10^{11}\unit{cm})^3$, at which
time the expansion is largely homologous in all cases considered, and the
internal energy is only a small fraction ($\mathord{<}0.5\%$) of the kinetic
energy.

Unlike the SPH simulations used for calculating the merging process, CASTRO is
unable to handle vacua.  The density of the ambient medium was set to
$10^{-1}\gcc$ at the beginning, and lowered by factors of 10 down to
$10^{-4}\gcc$ during remaps into bigger domains (to fill the volume not covered
by the old domain). For comparison, the highest densities at the end of the
simulations are $>20\gcc$ in all cases.  Low-density material is being swept up
at the explosion front, but appears to have no impact on the much denser bulk
of the ejecta inside, which expands homologously.  The total mass of the
ambient medium filling empty regions is always smaller than
$1.9\times10^{-3}\,\Msun$.

To increase the numerical stability during the shock breakout and to contain
detrimental effects on the time step, we employed a velocity cap of
$3\times10^9\cms\approx0.1c$. We kept track of the kinetic energy thus
subtracted from the system. The total subtracted energies are smaller than
$1.2\%$ of the total energy at the end of the simulations in all cases.

\section{Results}
\label{sec:results}

\renewcommand{\myscale}{.32}
\begin{figure*}[t]
\centering
\begin{tabular}{c@{}cc@{}cc}
\includegraphics[viewport=39 61 320 328,scale=\myscale]{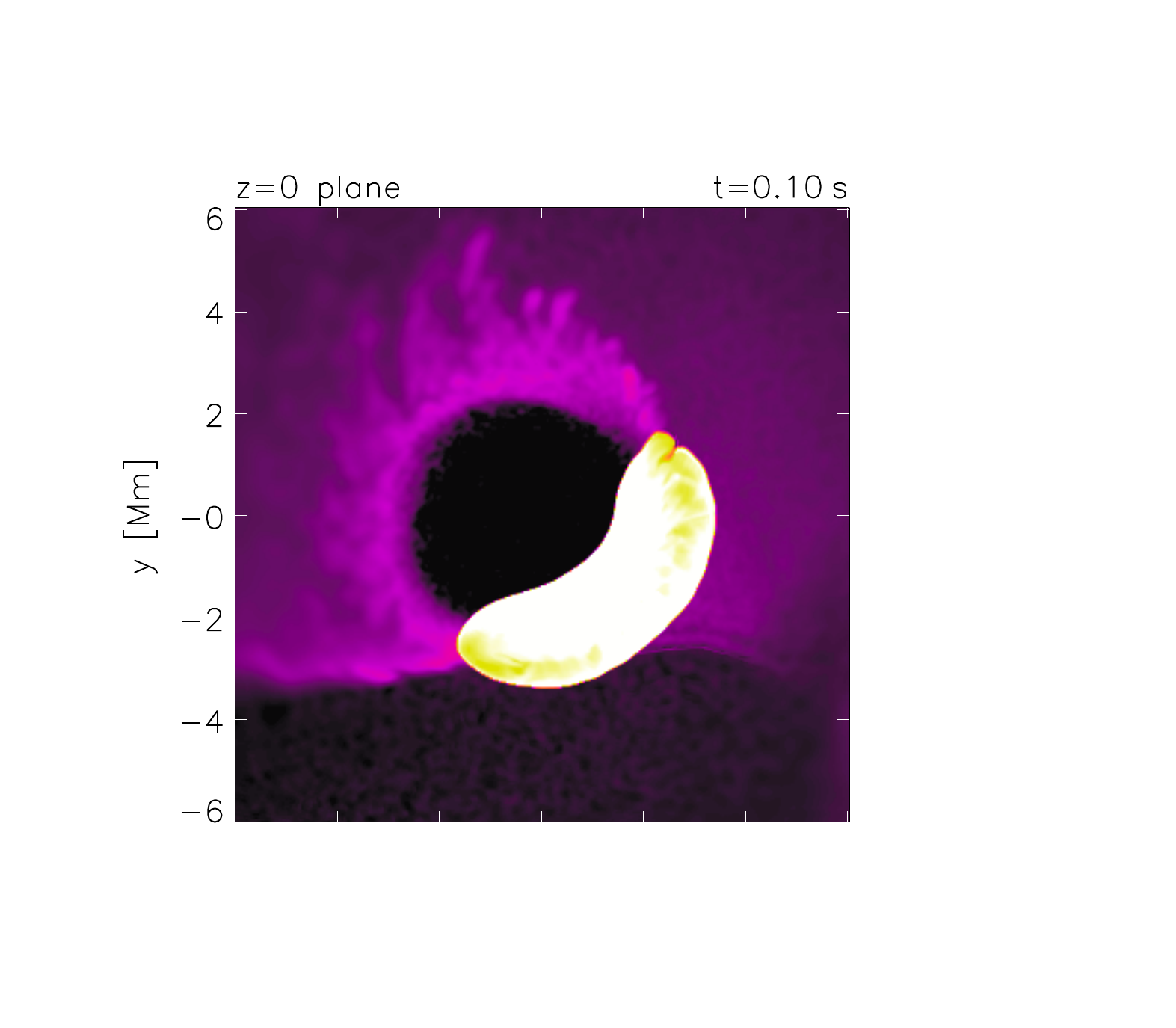} &
\includegraphics[viewport=39 61 320 328,scale=\myscale]{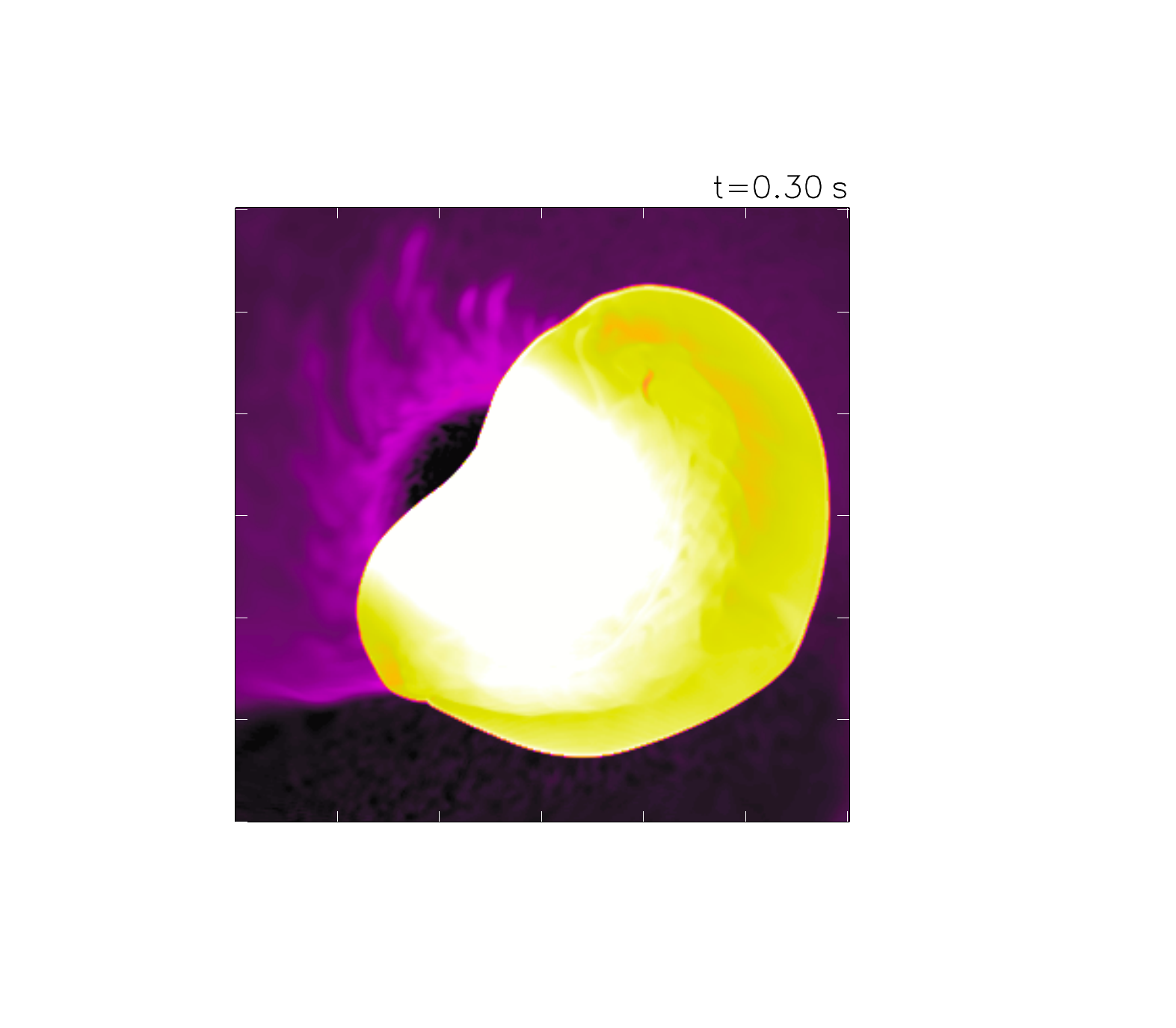} &
\includegraphics[viewport=39 61 320 328,scale=\myscale]{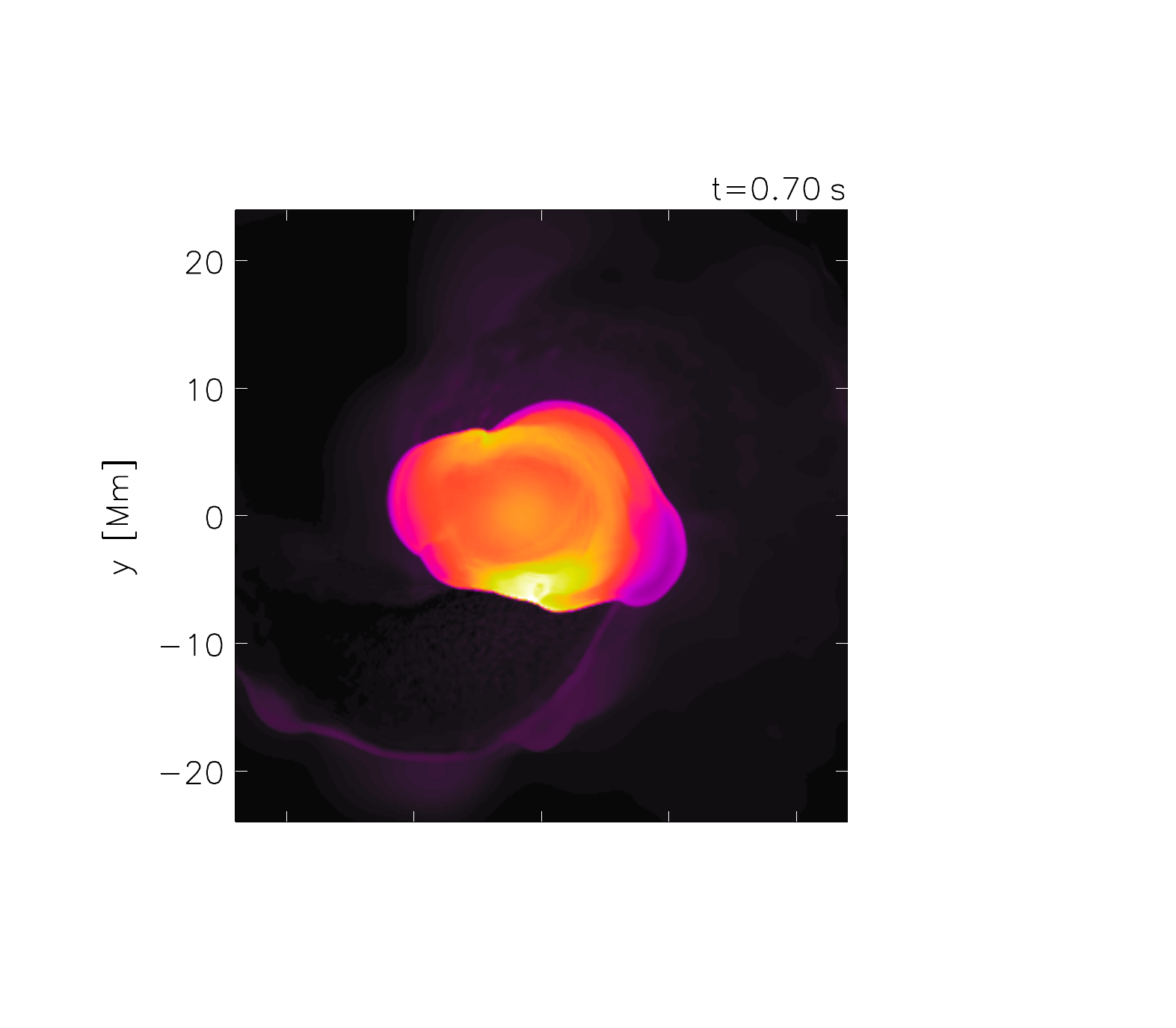} &
\includegraphics[viewport=39 61 320 328,scale=\myscale]{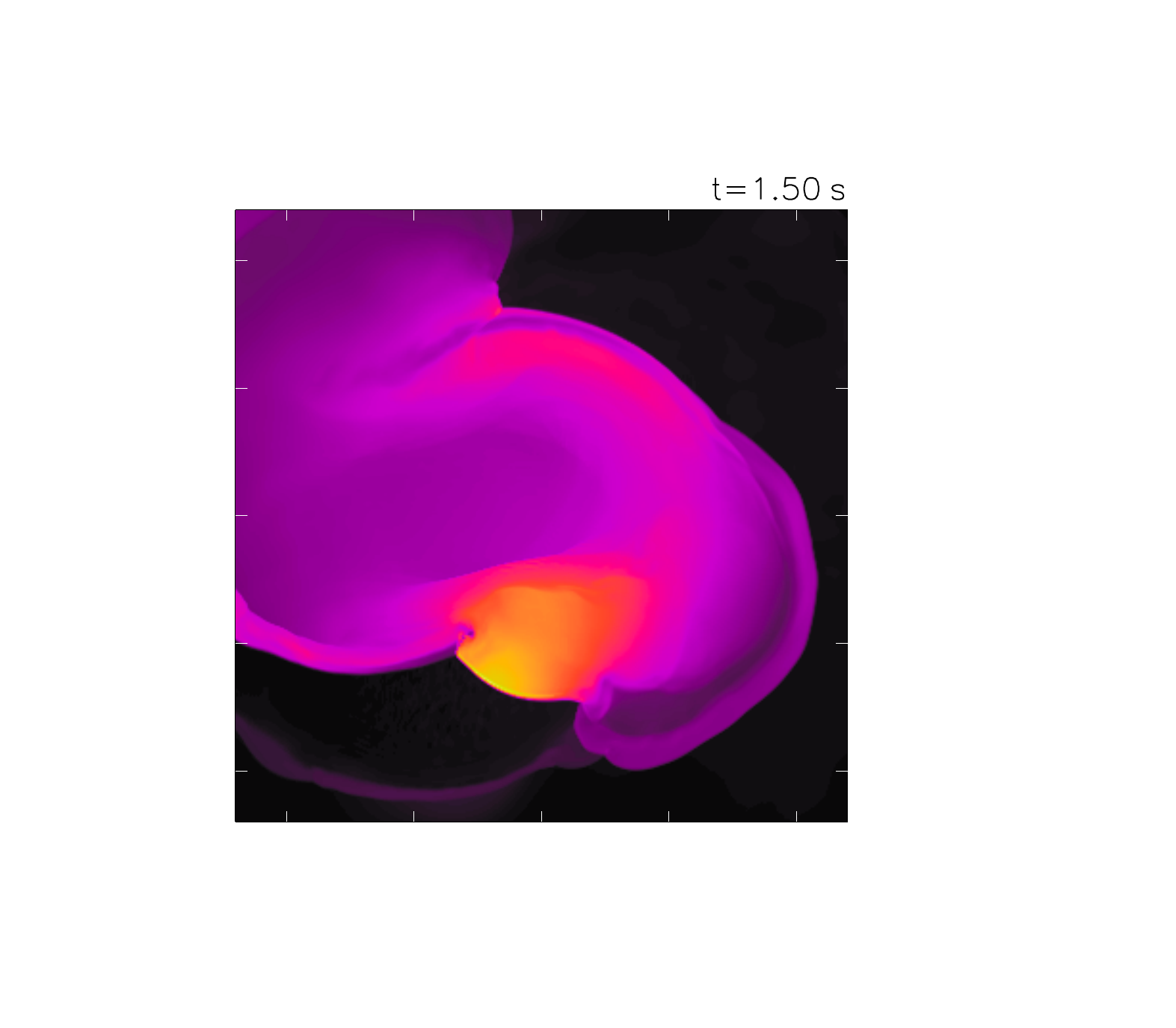} &
\includegraphics[viewport=39 61 380 328,scale=\myscale]{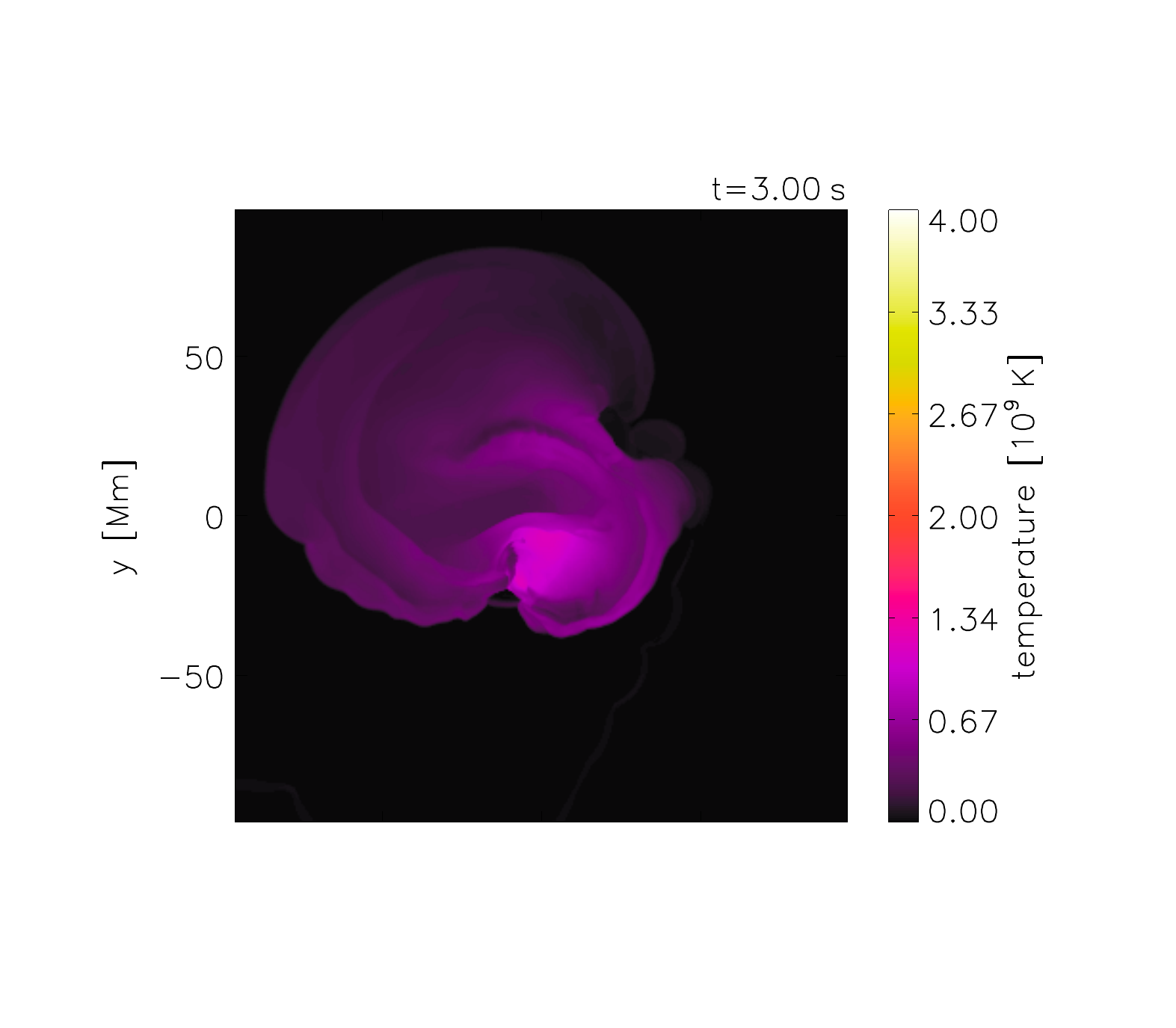} \\
\includegraphics[viewport=39 61 320 328,scale=\myscale]{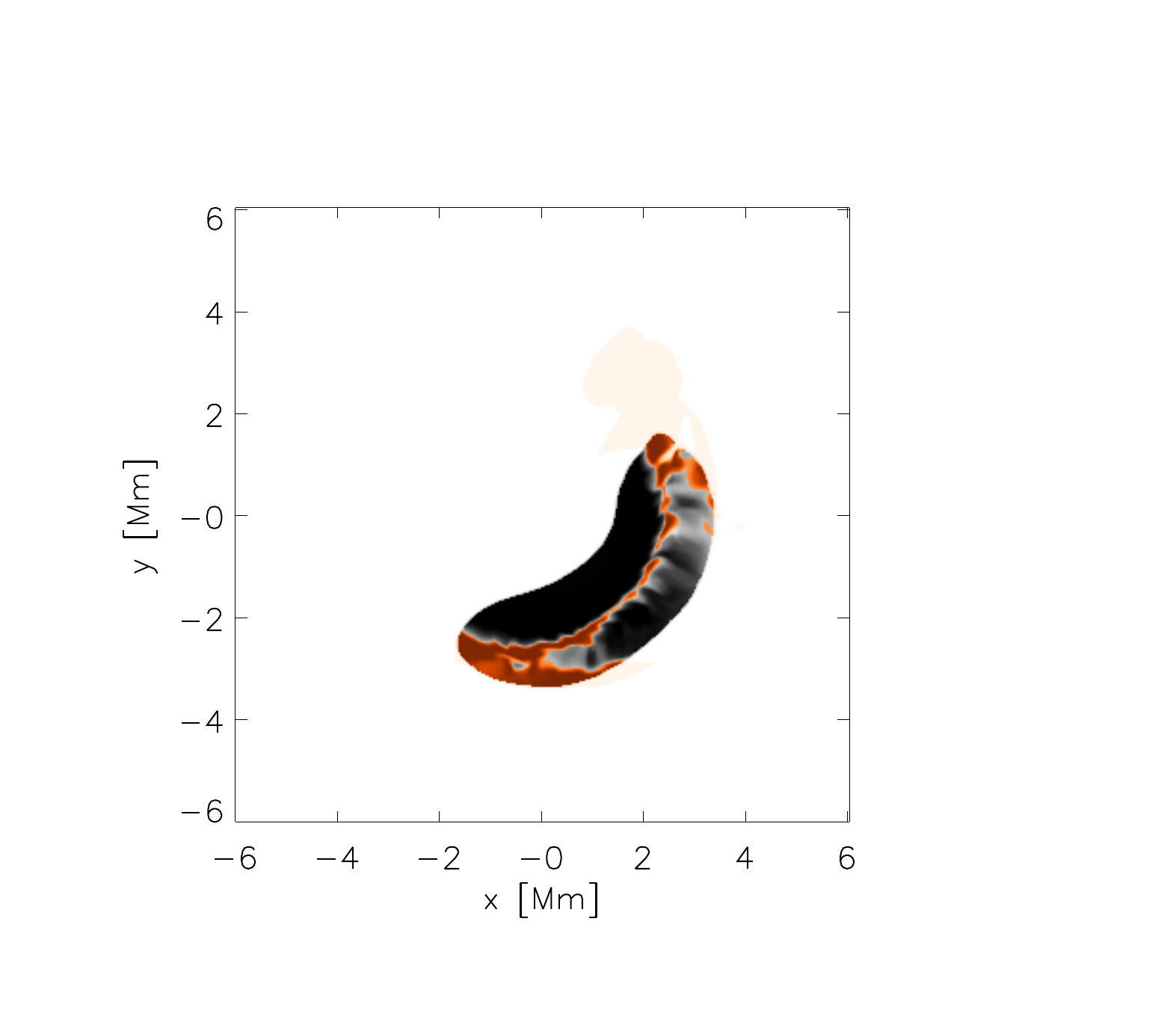} &
\includegraphics[viewport=39 61 320 328,scale=\myscale]{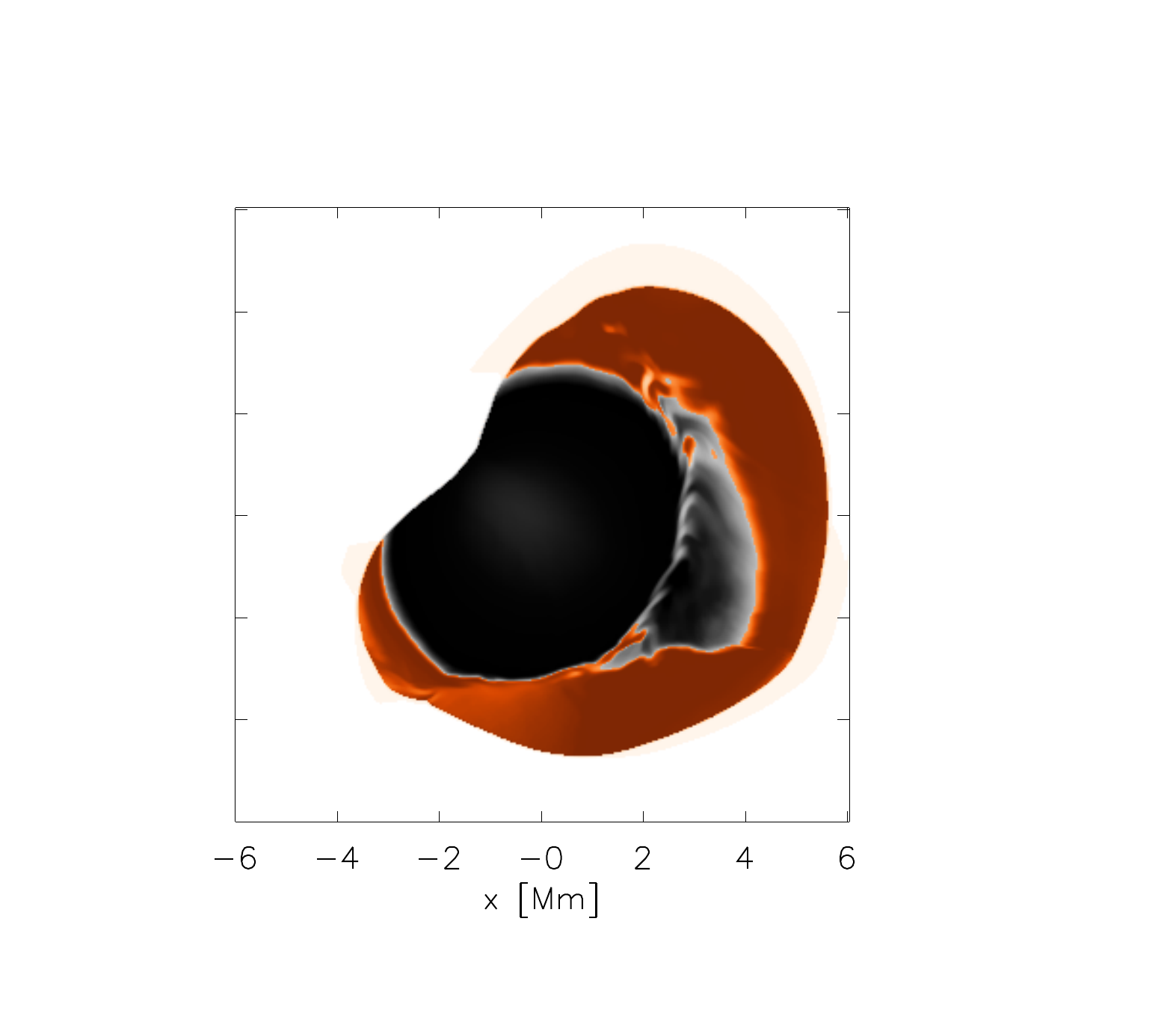} &
\includegraphics[viewport=39 61 320 328,scale=\myscale]{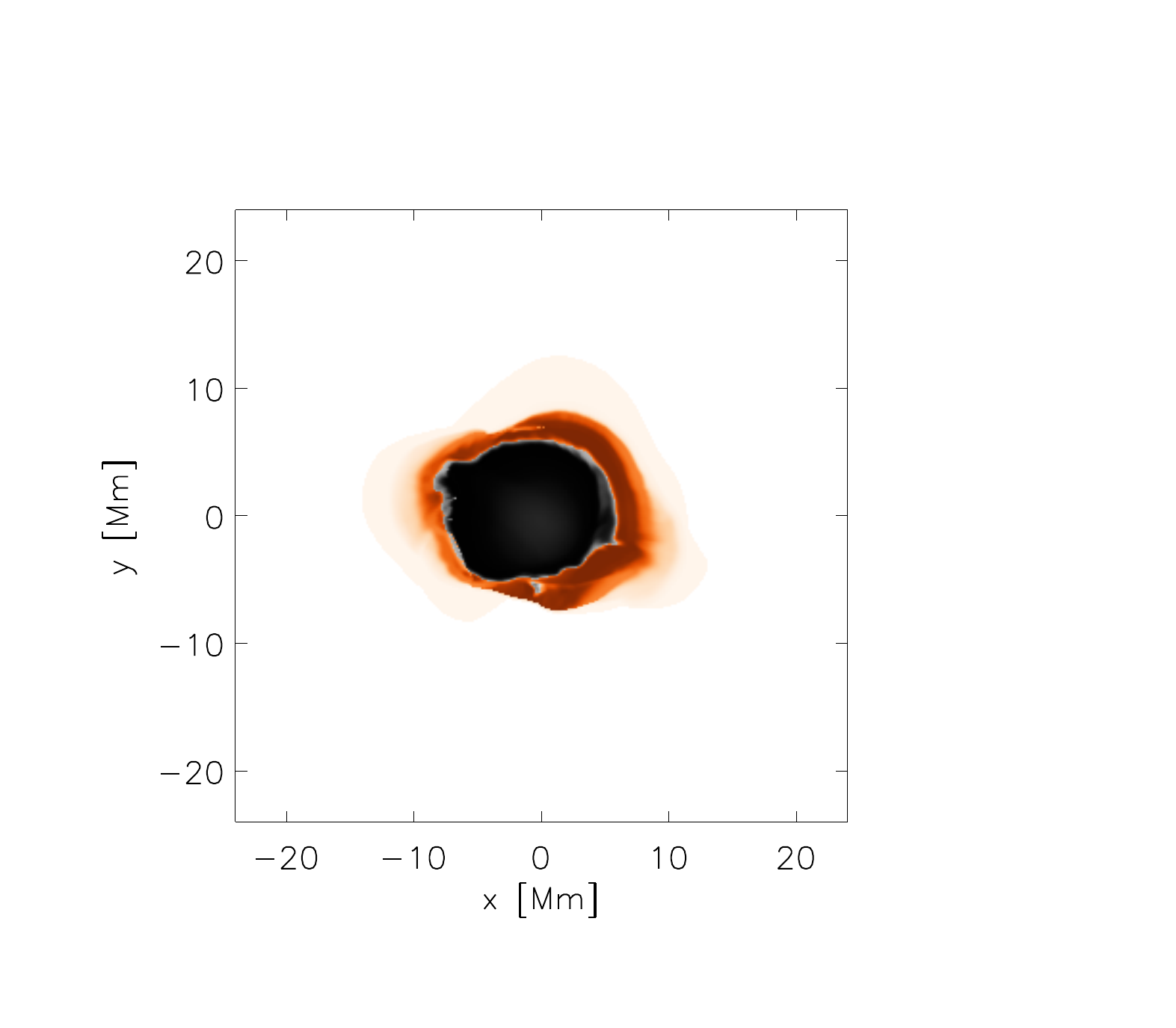} &
\includegraphics[viewport=39 61 320 328,scale=\myscale]{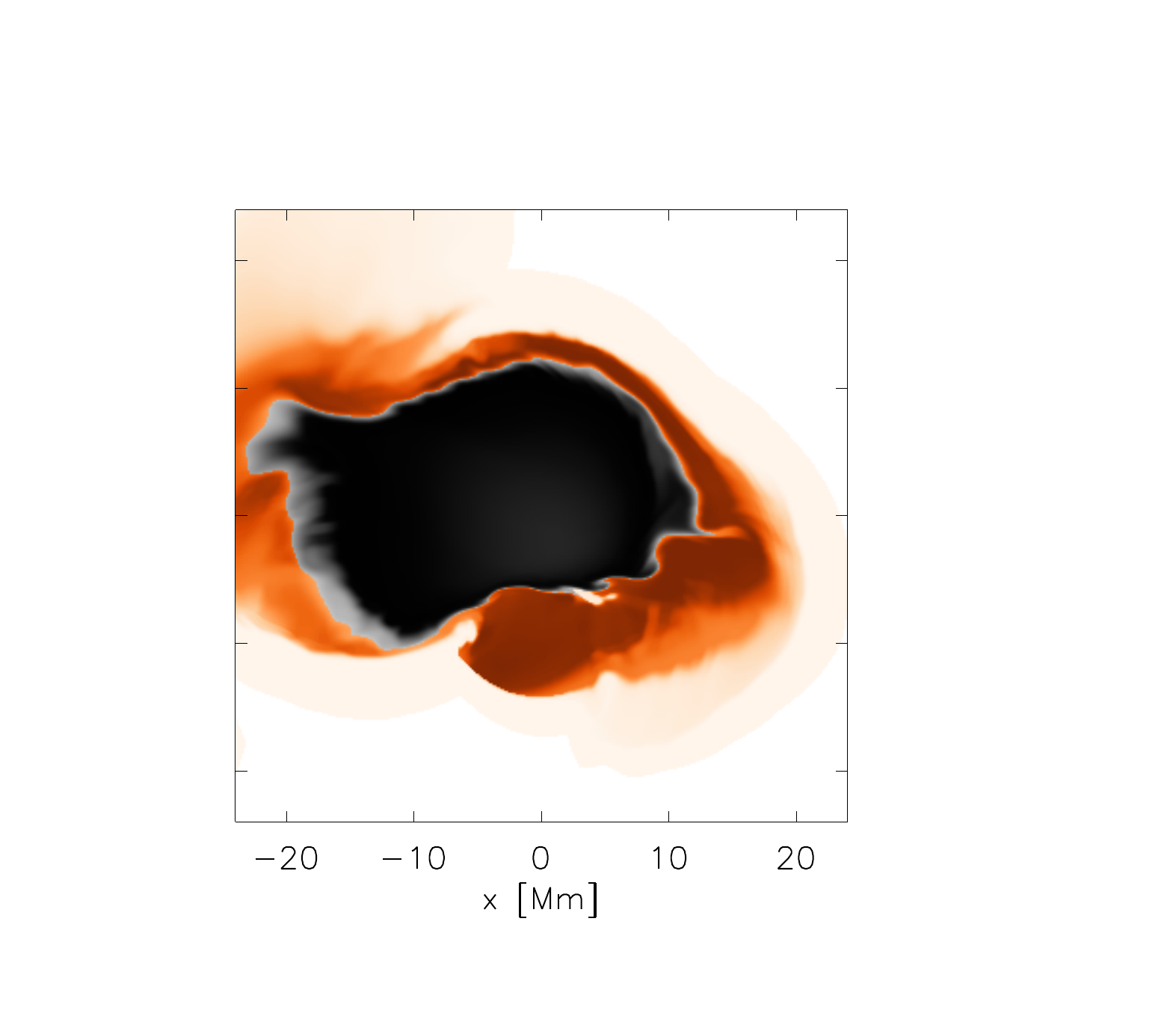} &
\includegraphics[viewport=39 61 380 328,scale=\myscale]{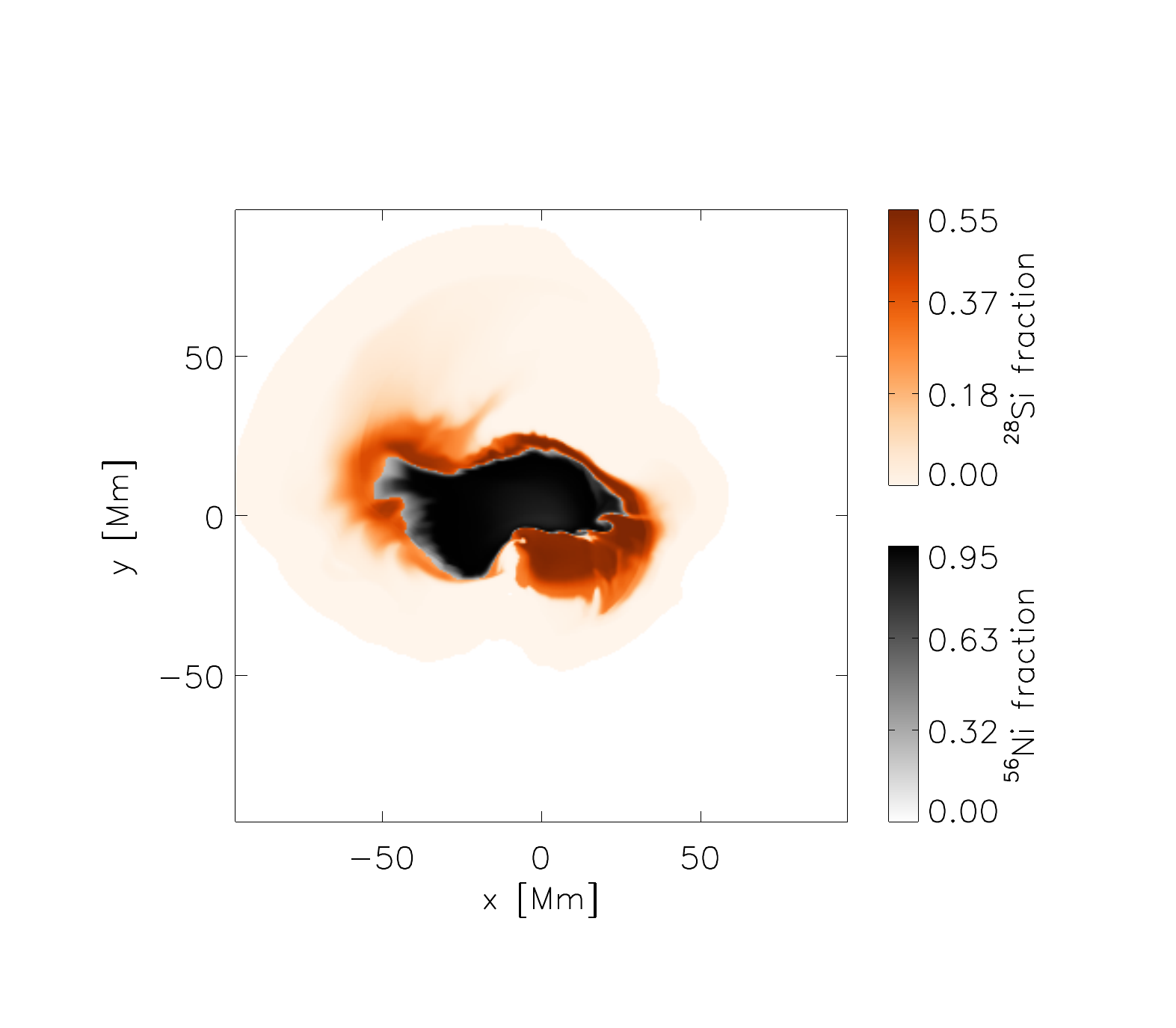}
\end{tabular}
\caption{Snapshots of the temperature (\textit{top row}) and the mass fractions
of \el{Ni}{56} and \el{Si}{28} (\textit{bottom row}, \el{Ni}{56} in black and
\el{Si}{28} in orange colors) in the orbital plane of model \mtwo.  For the
plots in the \textit{bottom row}, only the element with the respective larger
mass fraction is plotted at each pixel.}
\label{fig:expl_t0403}
\end{figure*}

\begin{figure}[t]
\includegraphics[width=\linewidth]{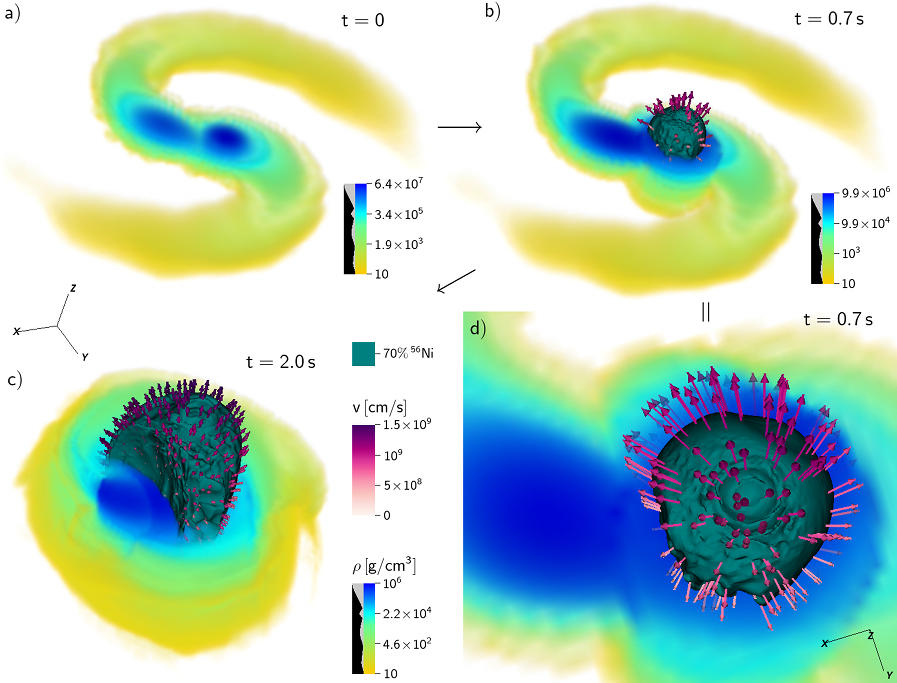}
\caption{Volume renderings of the density in model \mone right before the
detonation (panel a) and after the consumption of the primary star. The density
plot is clipped at the orbital plane, i.e., only half of the volume is shown.
The teal isosurface indicates the main pocket of \el{Ni}{56}.  The pink/magenta
arrows protruding from this surface represent the fluid velocity on the surface
(arrows are scaled by magnitude).  Panel (d) shows the same data as panel (b),
but at a different perspective and higher zoom level.  For a sense of scale:
the largest extent of the \el{Ni}{56} surface is approximately $10^9\unit{cm}$
in panels (b) and (d), and $4.4\times10^9\unit{cm}$ in panel (c).}
\label{fig:densvol_t0401}
\end{figure}

\begin{deluxetable}{r|rrr}
\tablecaption{Nucleosynthesis yields [$\Msun$] and total energies [erg]}
\tablehead{ model & \mtwo & \mone & \mnine }
\startdata
\el{He}{4} & $3.53\times10^{-3}$ & $6.05\times10^{-3}$ & $4.46\times10^{-3}$ \\
\el{C}{12} & $0.119$ & $4.42\times10^{-2}$ & $0.101$ \\
\el{N}{14} & $2.29\times10^{-13}$ & $3.36\times10^{-14}$ & $2.33\times10^{-13}$ \\
\el{O}{16} & $0.448$ & $0.311$ & $0.422$ \\
\el{Ne}{20} & $2.33\times10^{-2}$ & $1.06\times10^{-2}$ & $1.06\times10^{-2}$ \\
\el{Mg}{24} & $5.70\times10^{-2}$ & $4.32\times10^{-2}$ & $4.89\times10^{-2}$ \\
\el{Si}{28} & $0.381$ & $0.485$ & $0.389$ \\
\el{S}{32} & $0.138$ & $0.269$ & $0.155$ \\
\el{Ar}{36} & $1.22\times10^{-2}$ & $3.20\times10^{-2}$ & $1.82\times10^{-2}$ \\
\el{Ca}{40} & $6.78\times10^{-3}$ & $2.24\times10^{-2}$ & $1.41\times10^{-2}$ \\
\el{Ti}{44} & $6.21\times10^{-6}$ & $1.68\times10^{-5}$ & $1.32\times10^{-5}$ \\
\el{Cr}{48} & $1.93\times10^{-4}$ & $4.10\times10^{-4}$ & $3.51\times10^{-4}$ \\
\el{Fe}{52} & $5.22\times10^{-3}$ & $8.07\times10^{-3}$ & $7.93\times10^{-3}$ \\
\el{Fe}{54} & $7.45\times10^{-2}$ & $3.50\times10^{-2}$ & $1.78\times10^{-2}$ \\
\el{Ni}{56} & $0.999$ & $0.863$ & $0.580$ \\ \hline
$28<A<40$ & $0.537$ & $0.808$ & $0.576$ \\
$A\geq44$ & $1.079$ & $0.906$ & $0.606$ \\ \hline
$E_\text{begin}$ & $-6.39\times10^{50}$ & $-5.17\times10^{50}$ & $-3.08\times10^{50}$ \\
$E_\text{end}$ & $2.24\times10^{51}$ & $2.42\times10^{51}$ & $1.80\times10^{51}$ \\
$\Delta E$ & $2.88\times10^{51}$ & $2.94\times10^{51}$ & $2.11\times10^{51}$
\enddata
\label{tab:yields}
\end{deluxetable}

\renewcommand{\myscale}{.45}
\begin{figure}[t]
\raggedright
a)\\
\centering
\includegraphics[scale=\myscale]{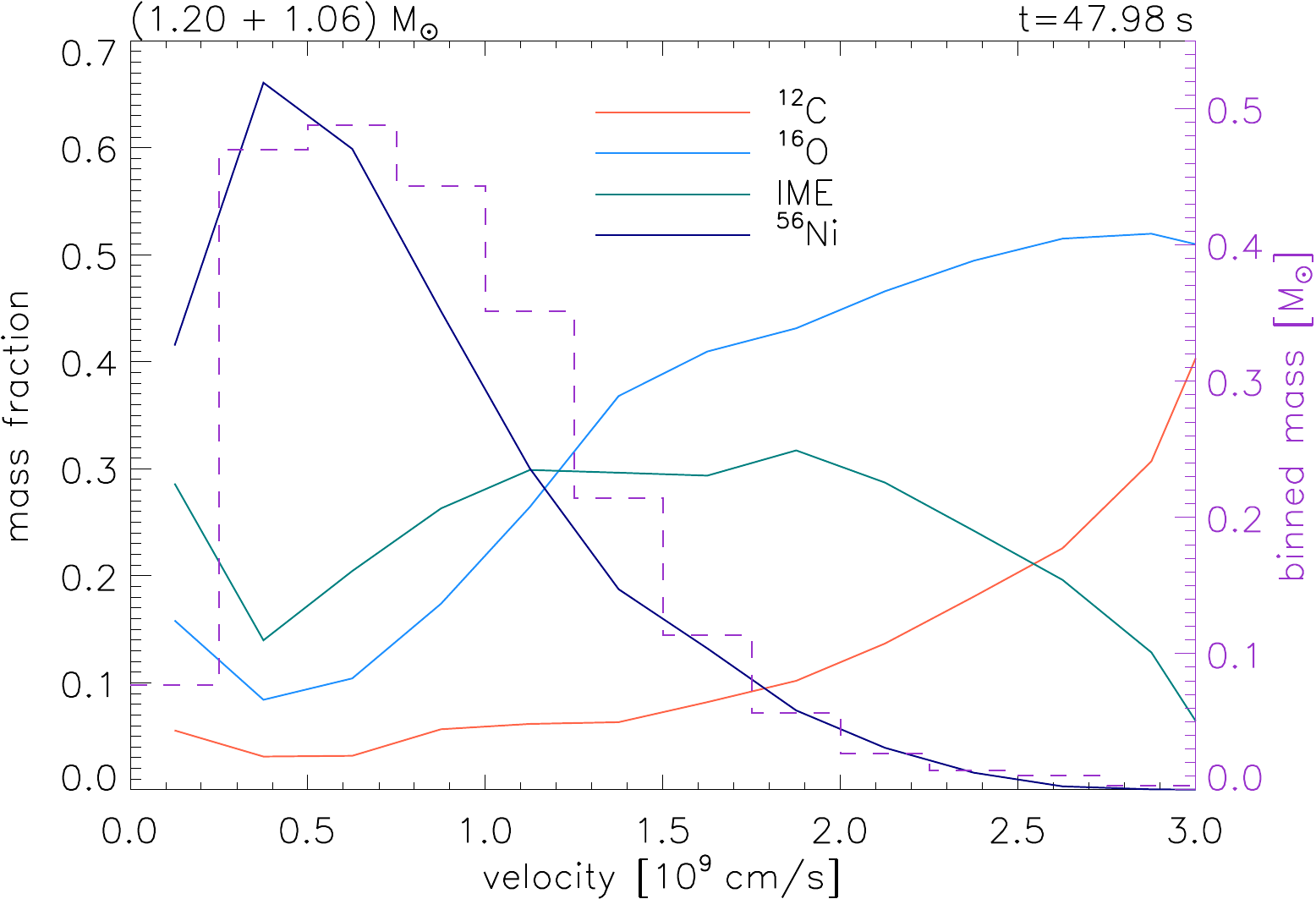} \\[\smallskipamount]
\raggedright
b)\\
\centering
\includegraphics[scale=\myscale]{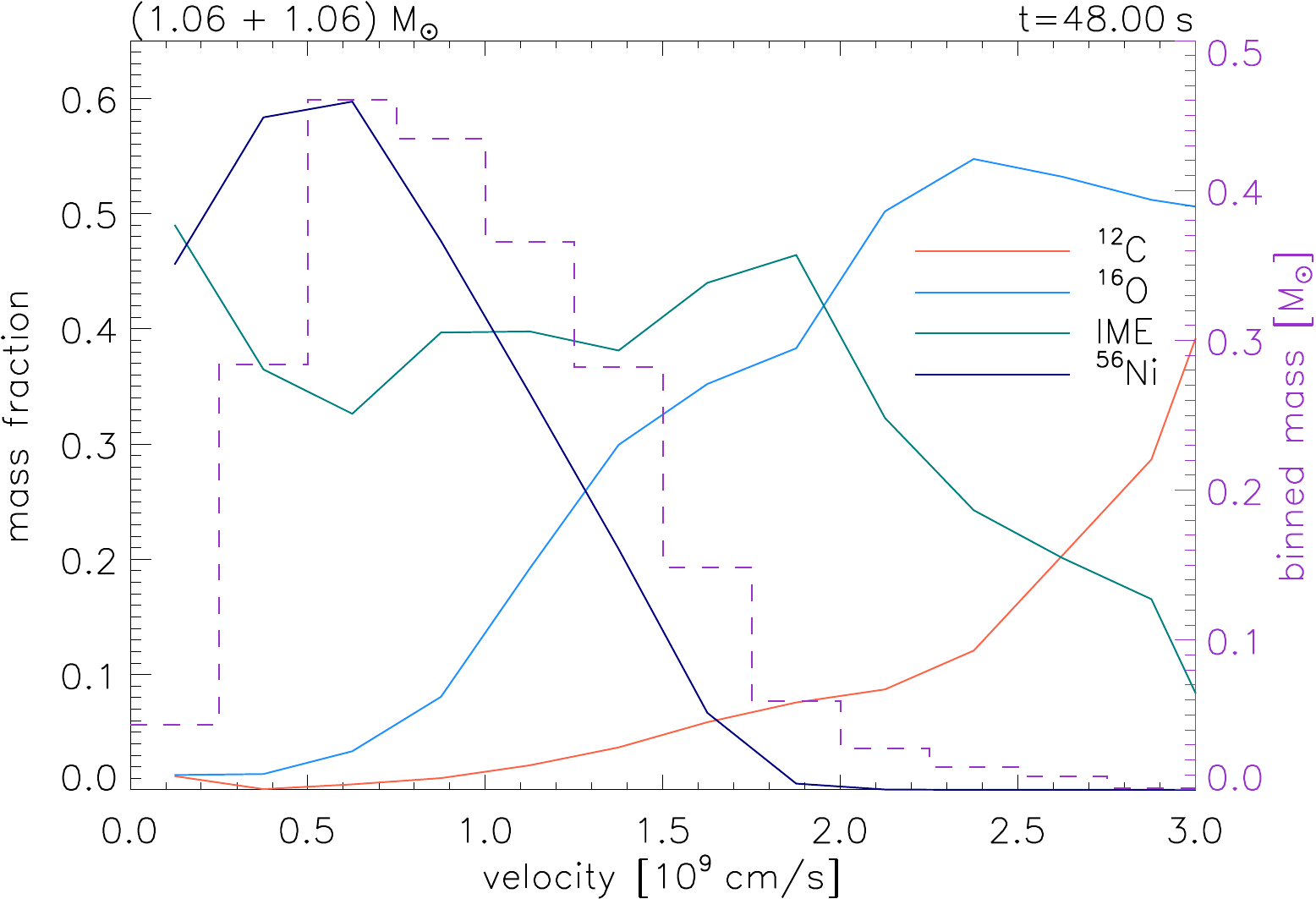} \\[\smallskipamount]
\raggedright
c)\\
\centering
\includegraphics[scale=\myscale]{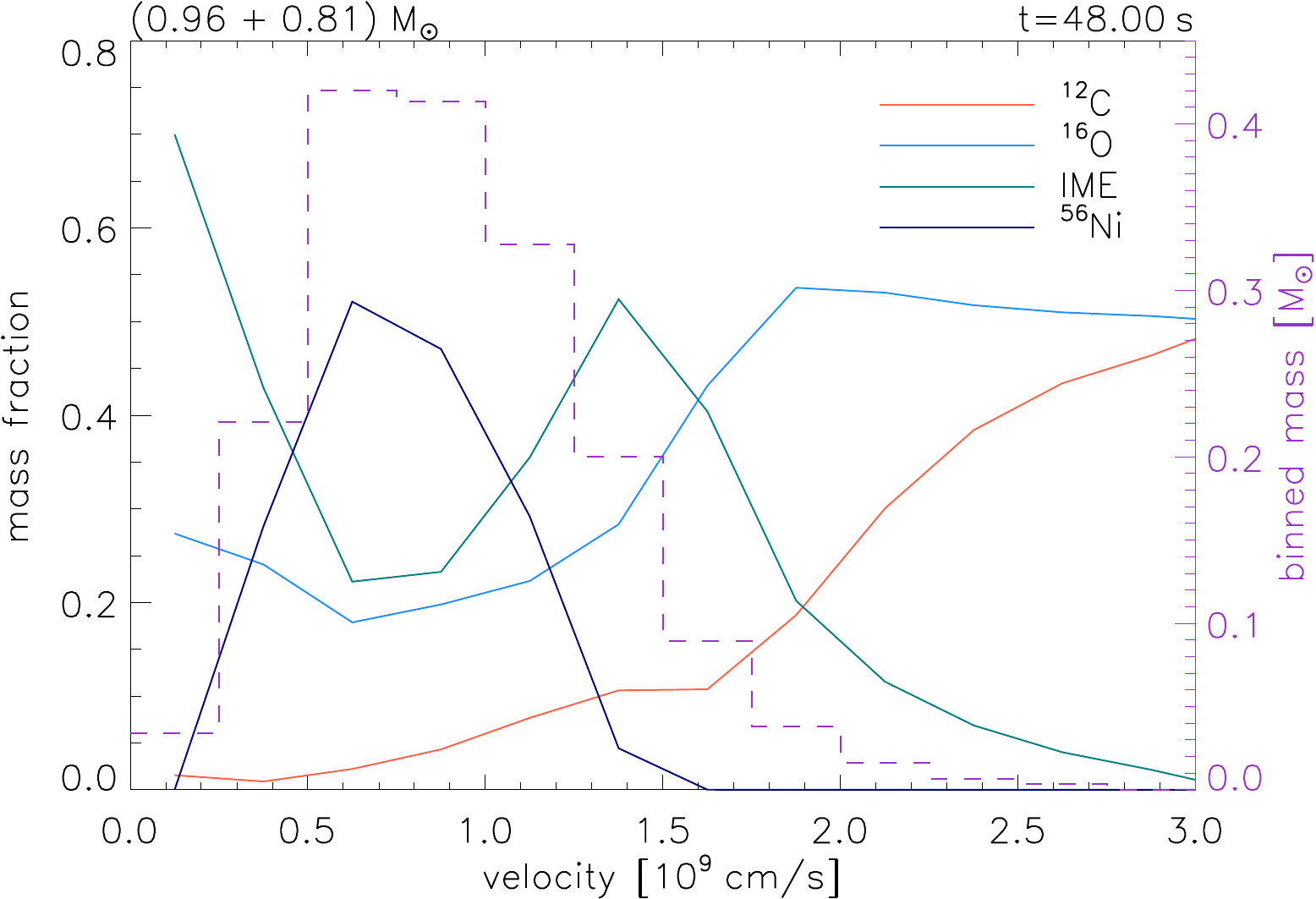} \\[\smallskipamount]
\raggedright
\caption{Mass fractions of different elements in bins of the ejecta velocity.
\el{Si}{28}, \el{S}{32}, \el{Ar}{36} and \el{Ca}{40} are grouped together as
``intermediate mass elements'' (IME, green line).  The total masses in the bins
used to calculate the fractions are shown by the dashed magenta line
(right-hand axis).}
\label{fig:massvel}
\end{figure}

\begin{figure*}[t]
\raggedright
a) $(1.20+1.06)\Msun$ \\
\centering
\includegraphics[width=.95\linewidth]{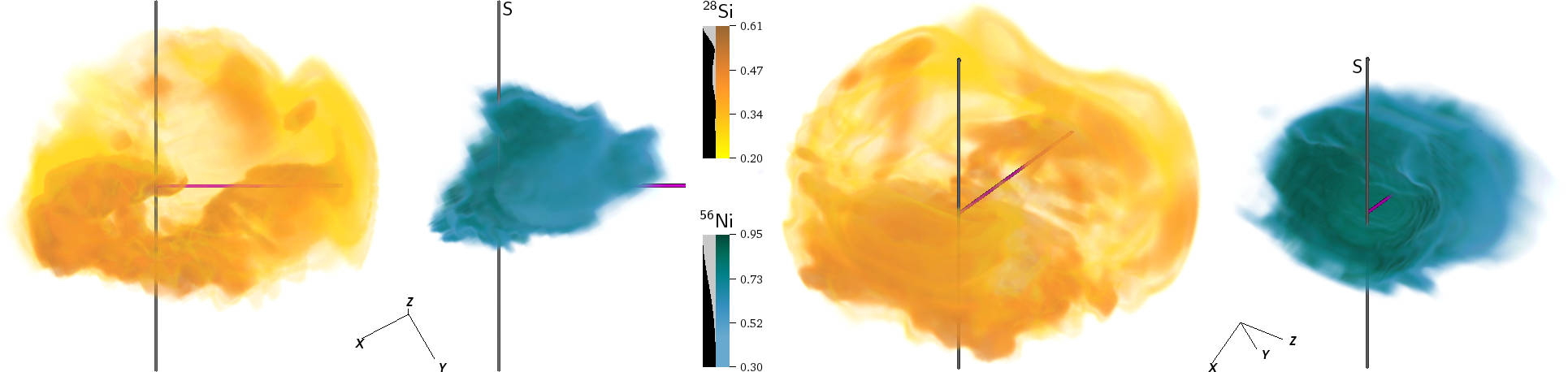} \\[\medskipamount]
\raggedright
b) $(1.06+1.06)\Msun$ \\
\centering
\includegraphics[width=.95\linewidth]{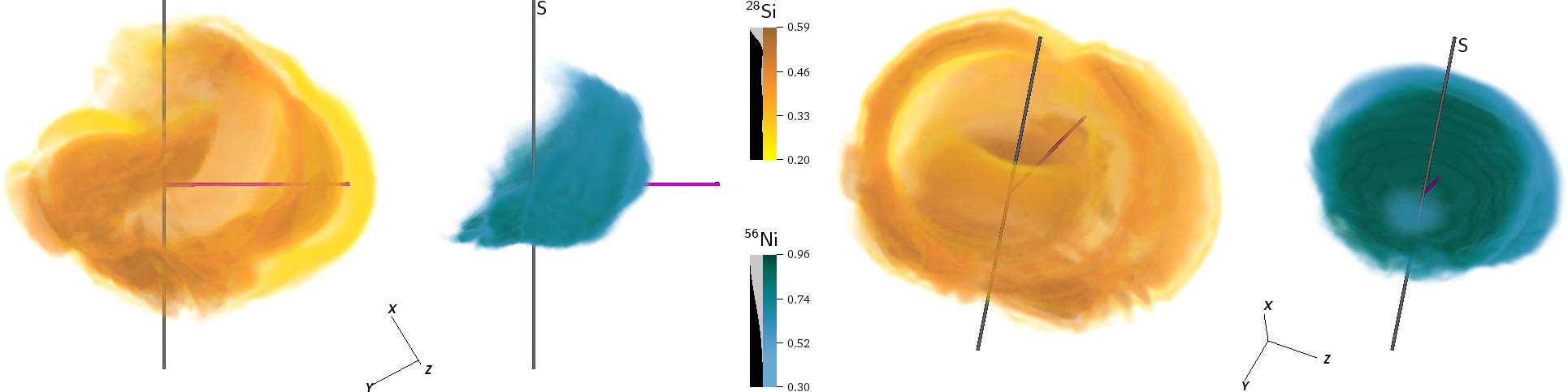} \\[\medskipamount]
\raggedright
c) $(0.96+0.81)\Msun$ \\
\centering
\includegraphics[width=.95\linewidth]{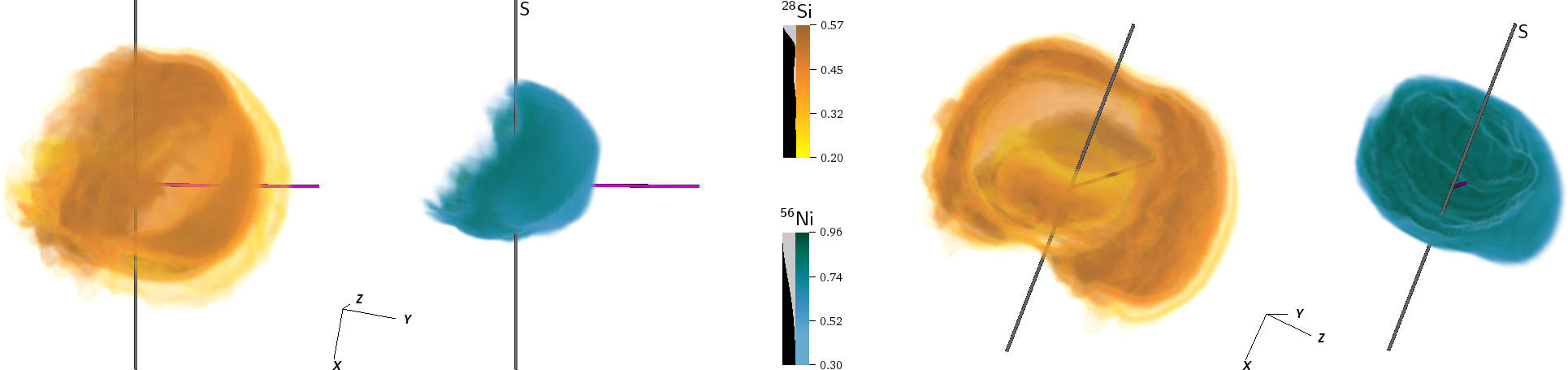}
\caption{Volume renderings of the element mass fractions of \el{Ni}{56} (teal
colors) and \el{Si}{28} (orange colors) in the homologously expanding ejecta,
48 seconds after the start of the detonation, at two different perspectives,
respectively.  The gray and magenta rods both lie in the orbital plane,
intersecting at the origin (cf.  Figure~\ref{fig:initrho}): the gray rod
represents the line connecting the centers of the two WDs when the
detonation sets off, and the magenta rod points in orthogonal direction away
from the tidal tail of the primary.  The side at which the secondary was is
marked with an ``S''.  The perspective to the left corresponds to a face-on
view of the orbital plane.  To give a sense of scale: the distance from the
origin to the respective outer tips of the rods is $10^{11}\unit{cm}$.  The
homologous expansion velocity at this distance is approximately
$2.15\times10^9\cms$ in all cases shown.}
\label{fig:ejecta}
\end{figure*}

\renewcommand{\myscale}{.25}
\begin{figure*}[t]
\centering
\begin{tabular}{lll}
a) $(1.20+1.06)\Msun$ &
b) $(1.06+1.06)\Msun$ &
c) $(0.96+0.81)\Msun$ \\
\includegraphics[scale=\myscale]{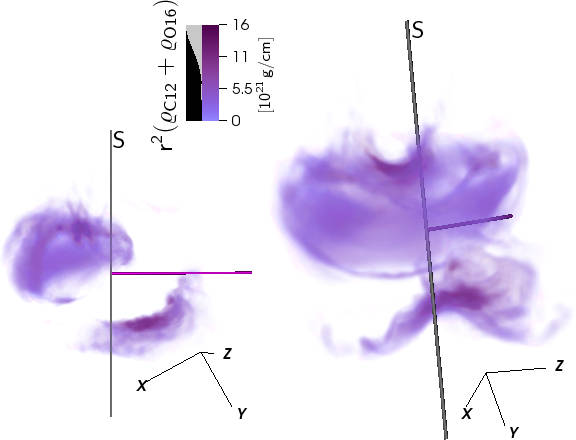} &
\includegraphics[scale=\myscale]{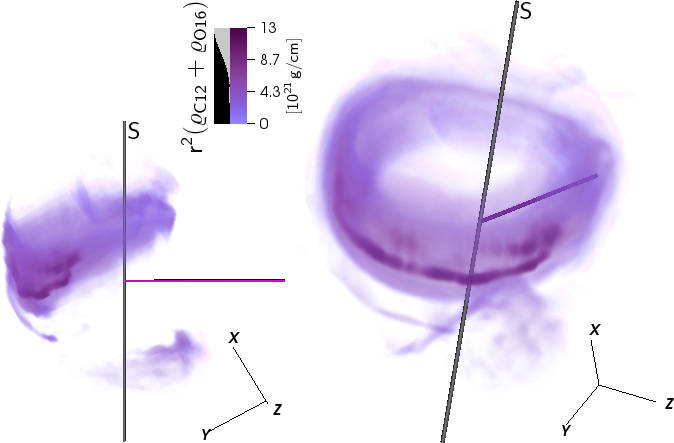} &
\includegraphics[scale=\myscale]{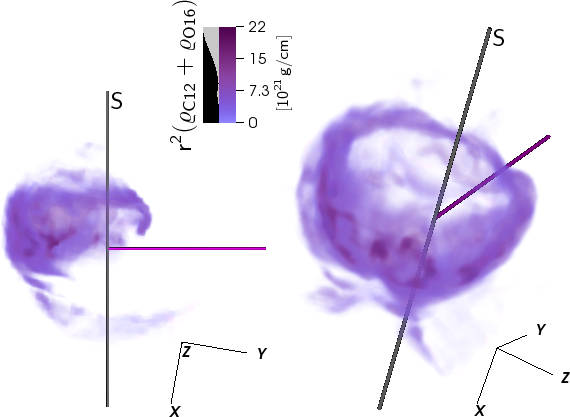}
\end{tabular}
\caption{Volume renderings of the ejecta of \el{C}{12} and \el{O}{16} (unburnt
fuel) in the homologously expanding ejecta, 48 seconds after the start of the
detonation, at two different perspectives, respectively.  The meaning of the
rods is the same as in Figure~\ref{fig:ejecta}, but for practical reasons, we
here plot the density times the square of the distance to the origin (thus
compensating for the increase of scales due to the expansion) instead of the
mass fraction.}
\label{fig:ejecCO}
\end{figure*}

\begin{figure}[t]
\includegraphics[width=.9\linewidth]{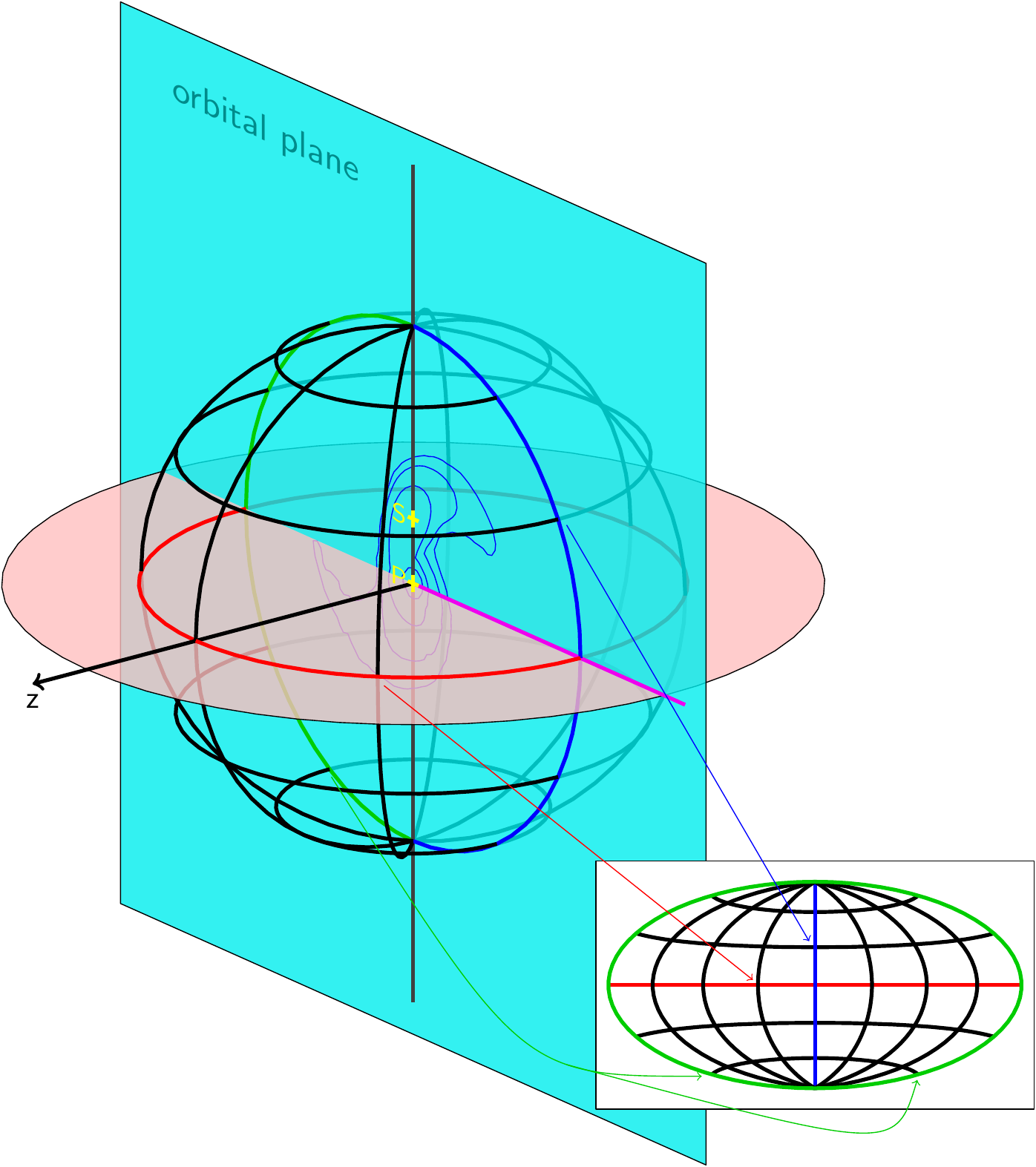}
\caption{Sketch of the map projection used in Figure~\ref{fig:sterad}.  The cyan
plane, on which density contours of a merger are drawn in blue, corresponds to
the orbital plane shown in Figure~\ref{fig:initrho}.  Yellow crosses represent
the centers of the merging stars.  The gray and magenta lines correspond to the
same-colored rods in Figures~\ref{fig:ejecta} and~\ref{fig:ejecCO}. In the map
projection, they correspond to the polar axis and the direction of the prime
meridian in the equatorial plane (shaded pink in the sketch), respectively.}
\label{fig:projection}
\end{figure}

\renewcommand{\myscale}{.40}
\begin{figure*}[p]
\centering
\begin{tabular}{lll}
a) & & \\
\includegraphics[scale=\myscale]{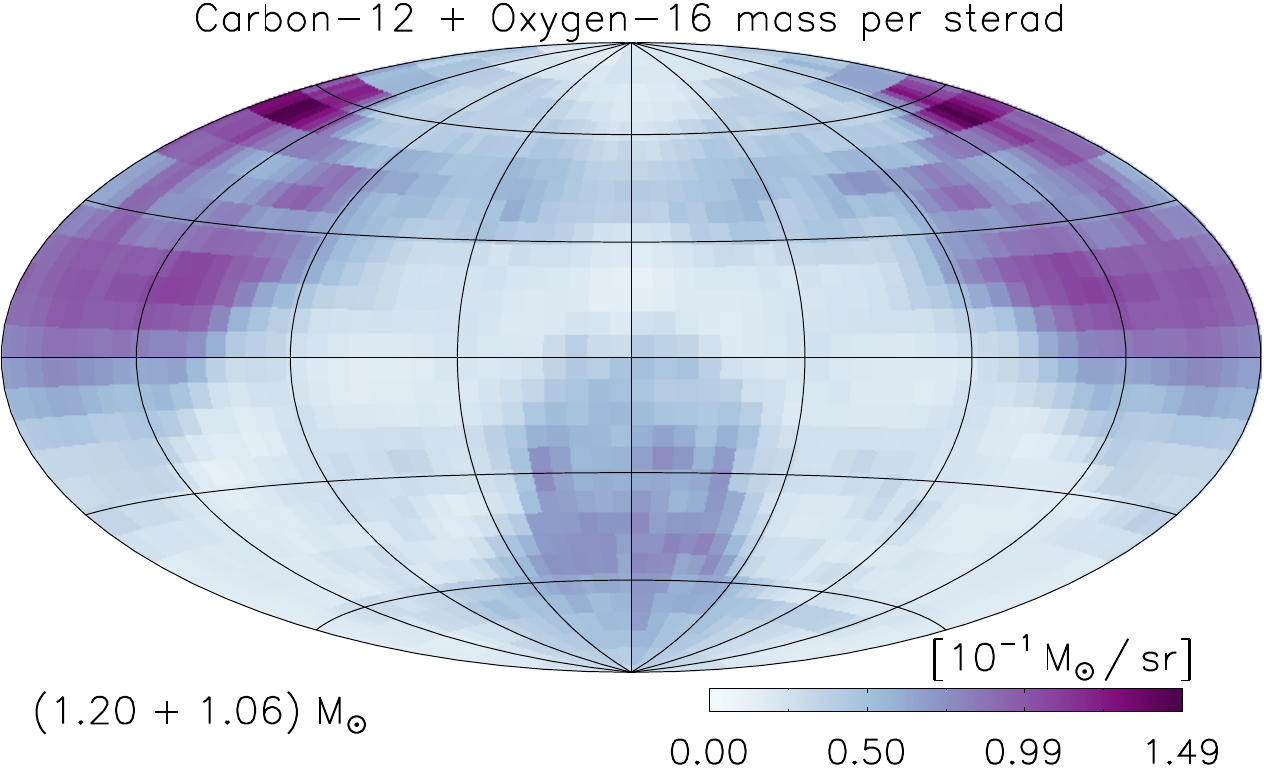} &
\includegraphics[scale=\myscale]{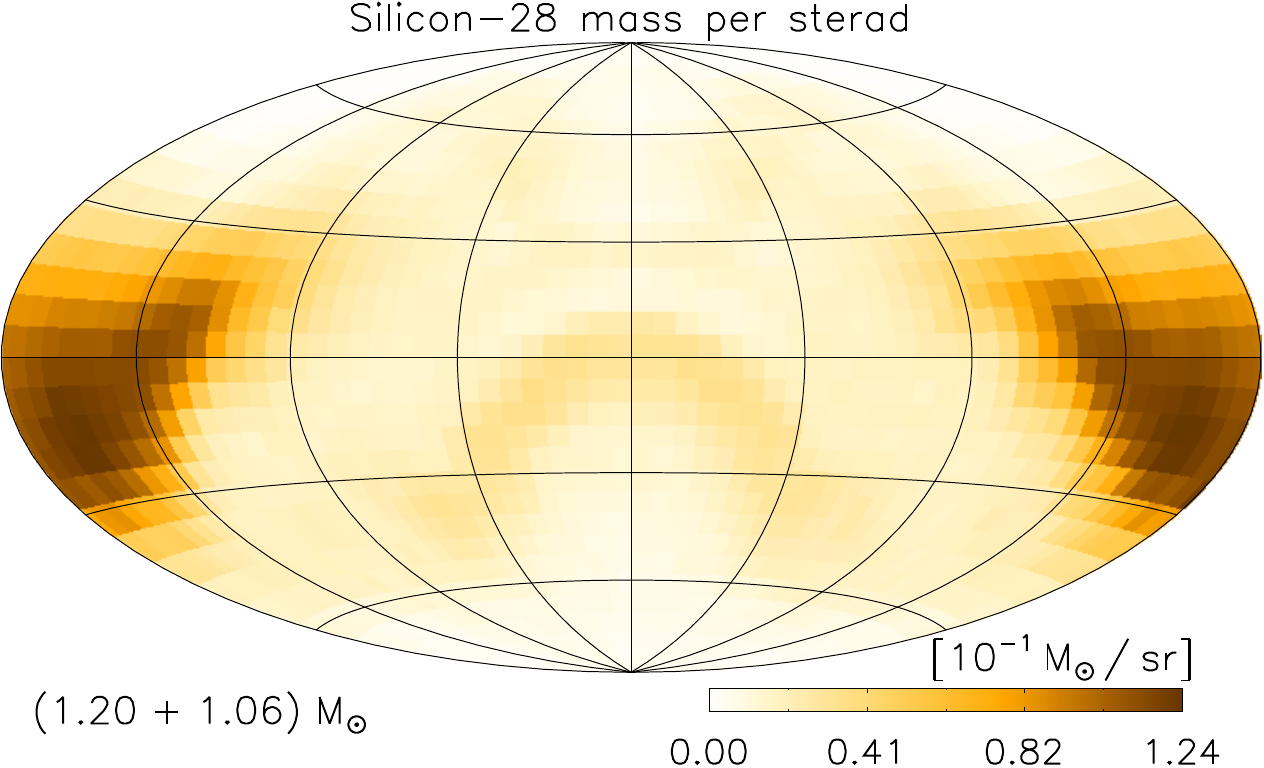} &
\includegraphics[scale=\myscale]{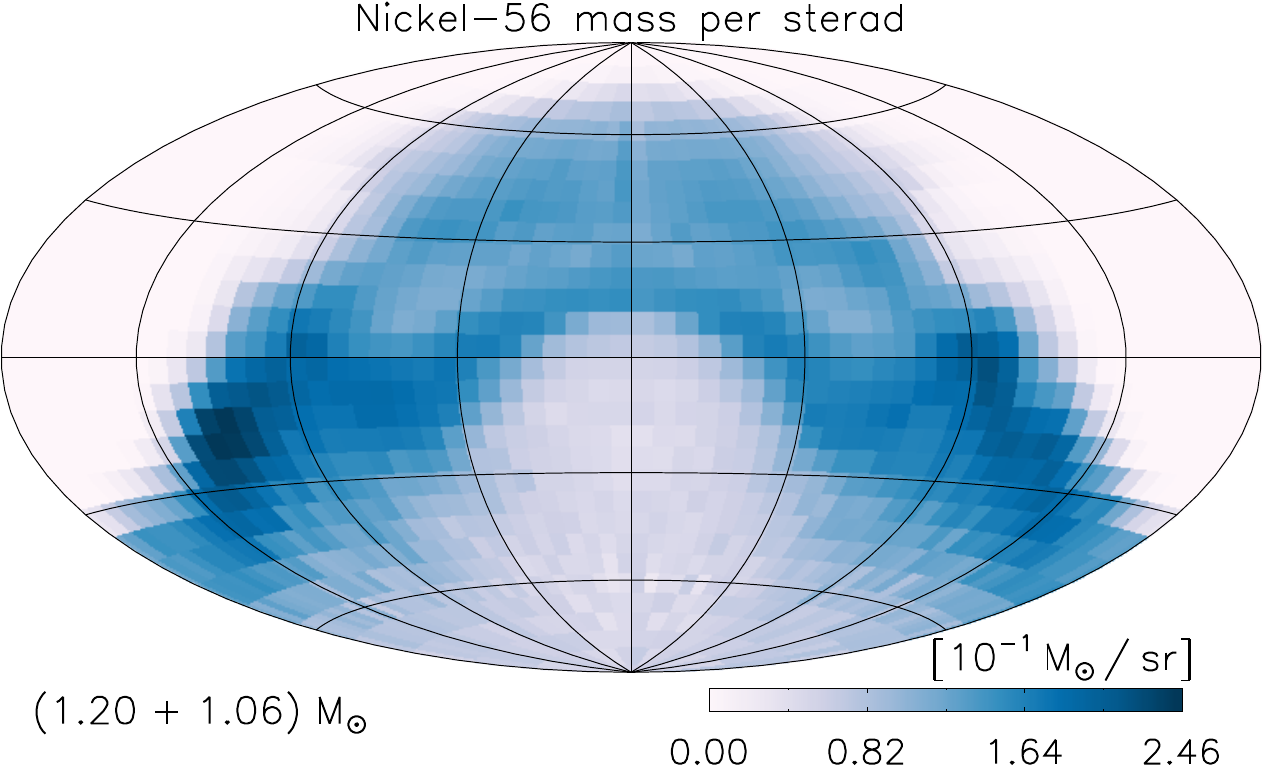} \\
\includegraphics[scale=\myscale]{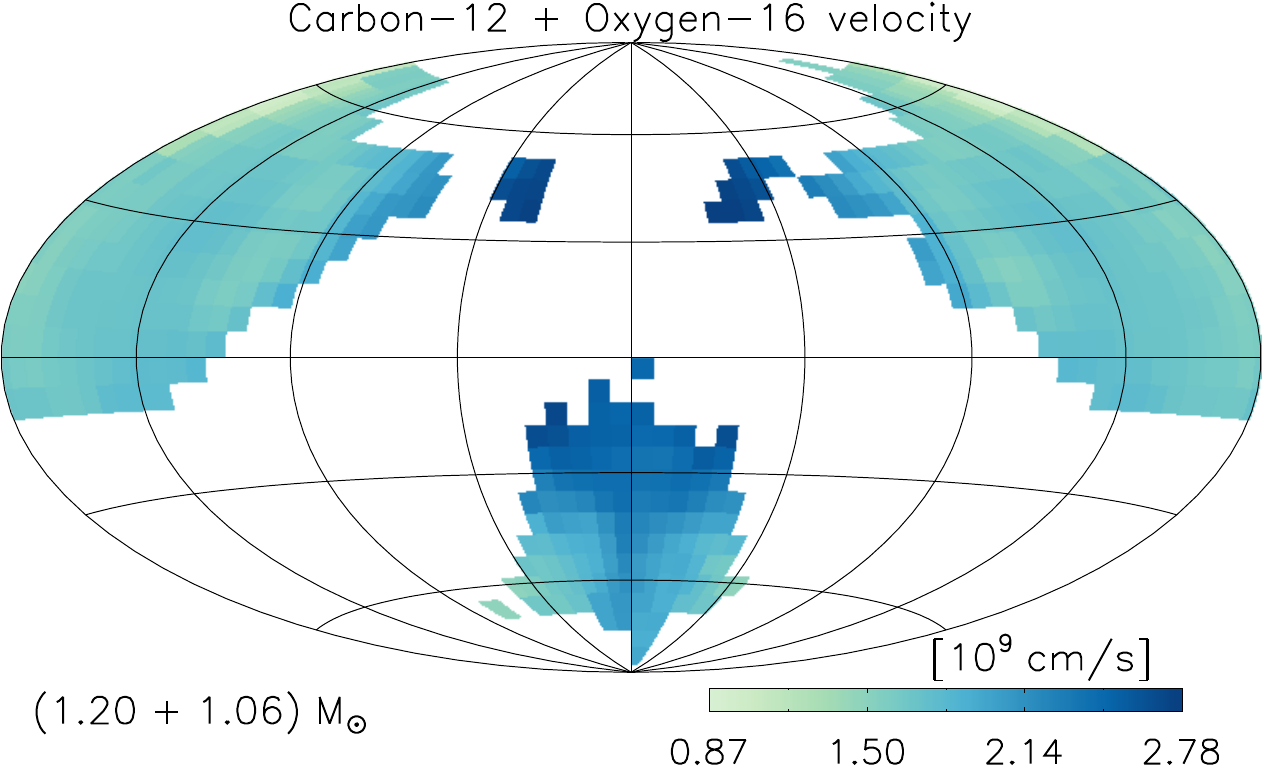} &
\includegraphics[scale=\myscale]{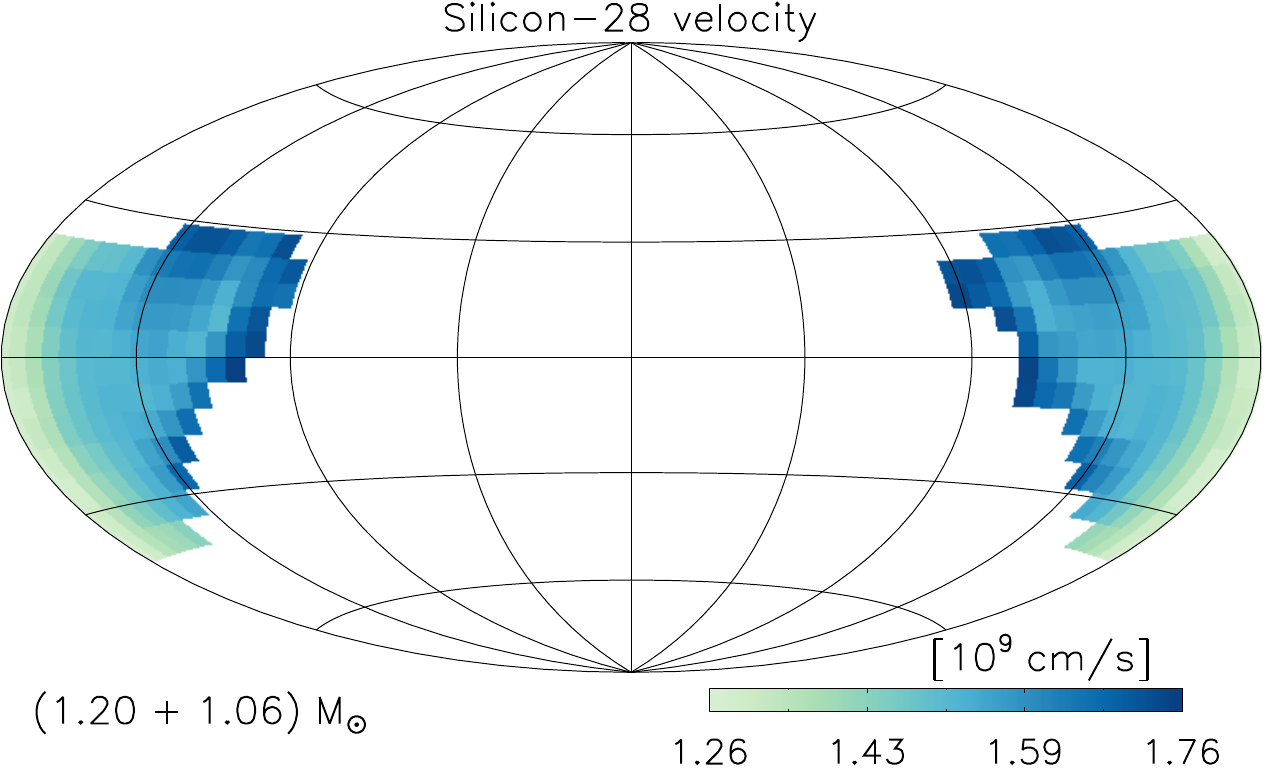} &
\includegraphics[scale=\myscale]{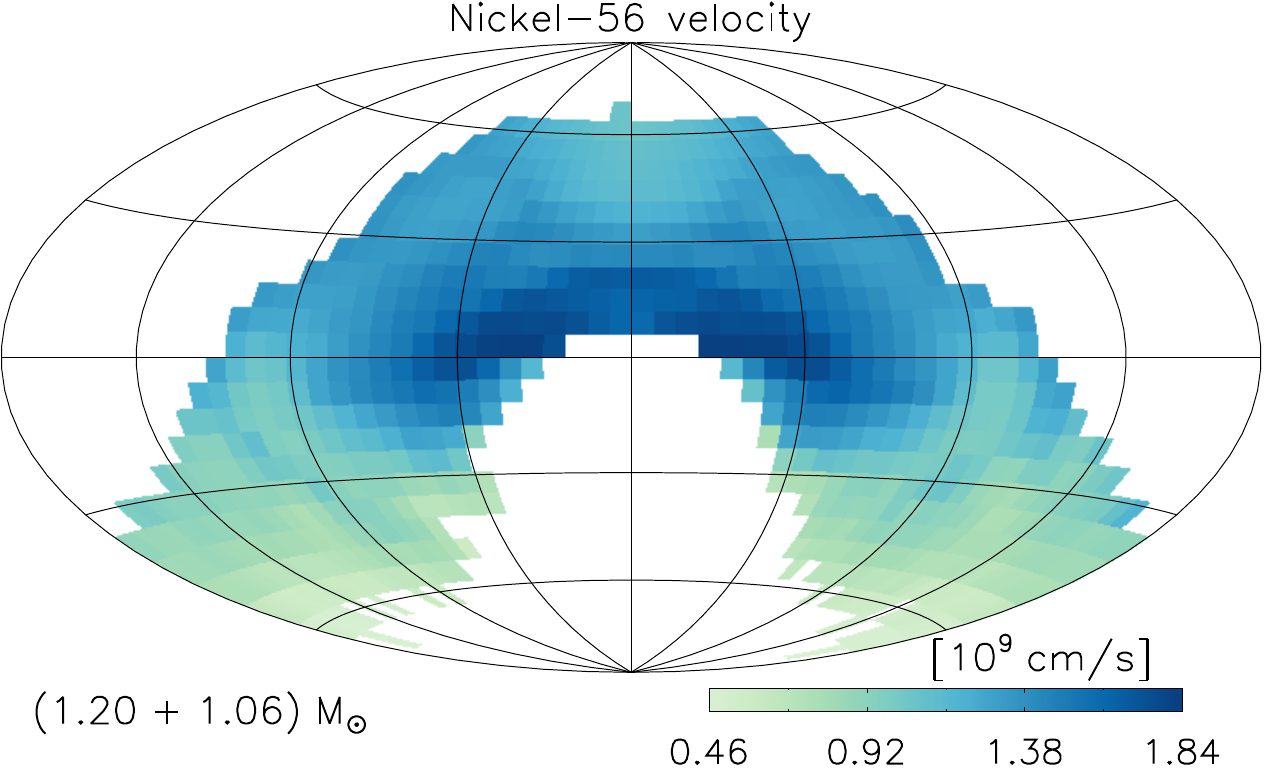} \\[\bigskipamount]
b) & &\\
\includegraphics[scale=\myscale]{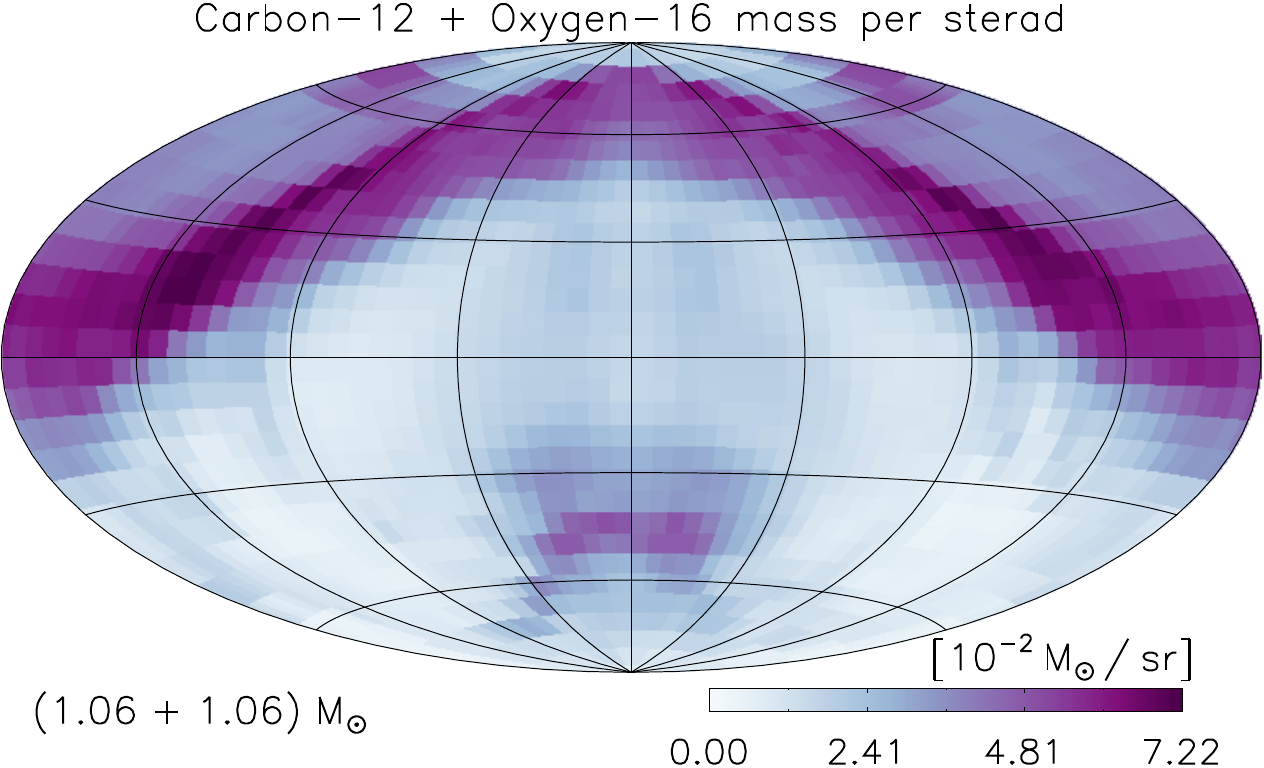} &
\includegraphics[scale=\myscale]{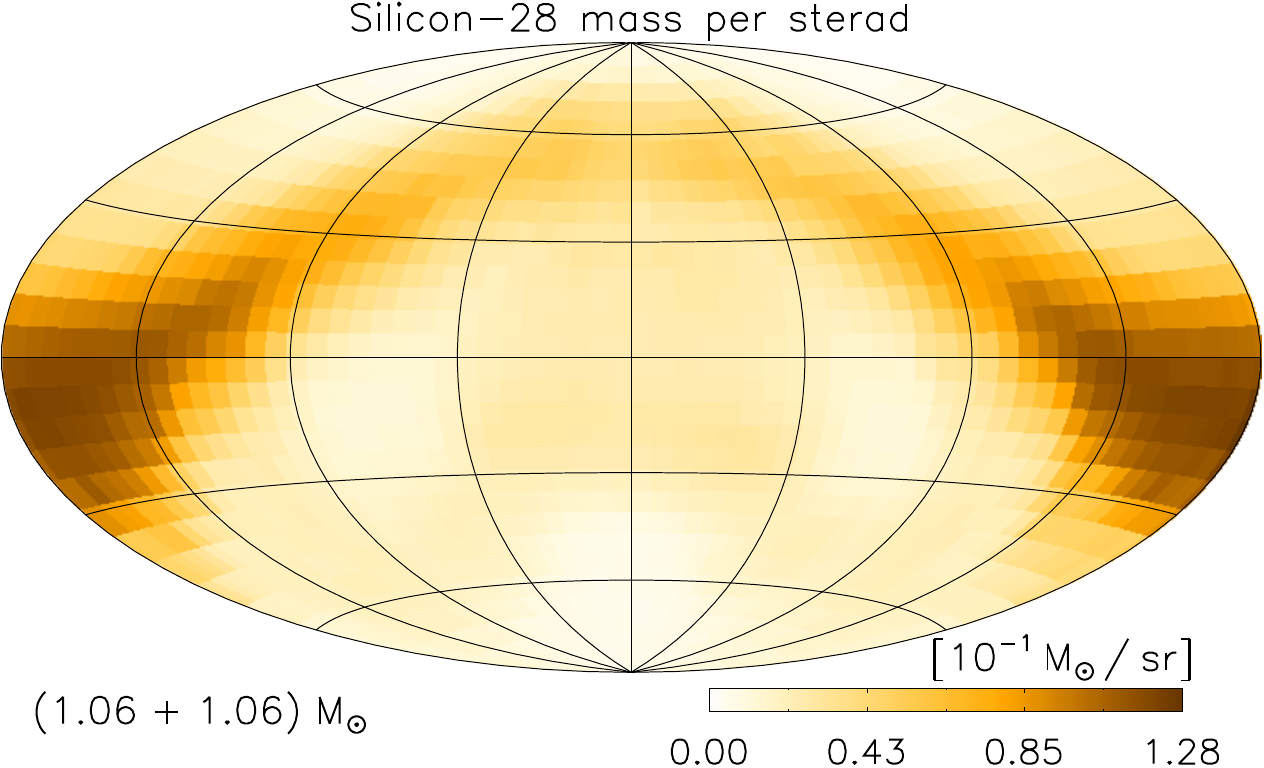} &
\includegraphics[scale=\myscale]{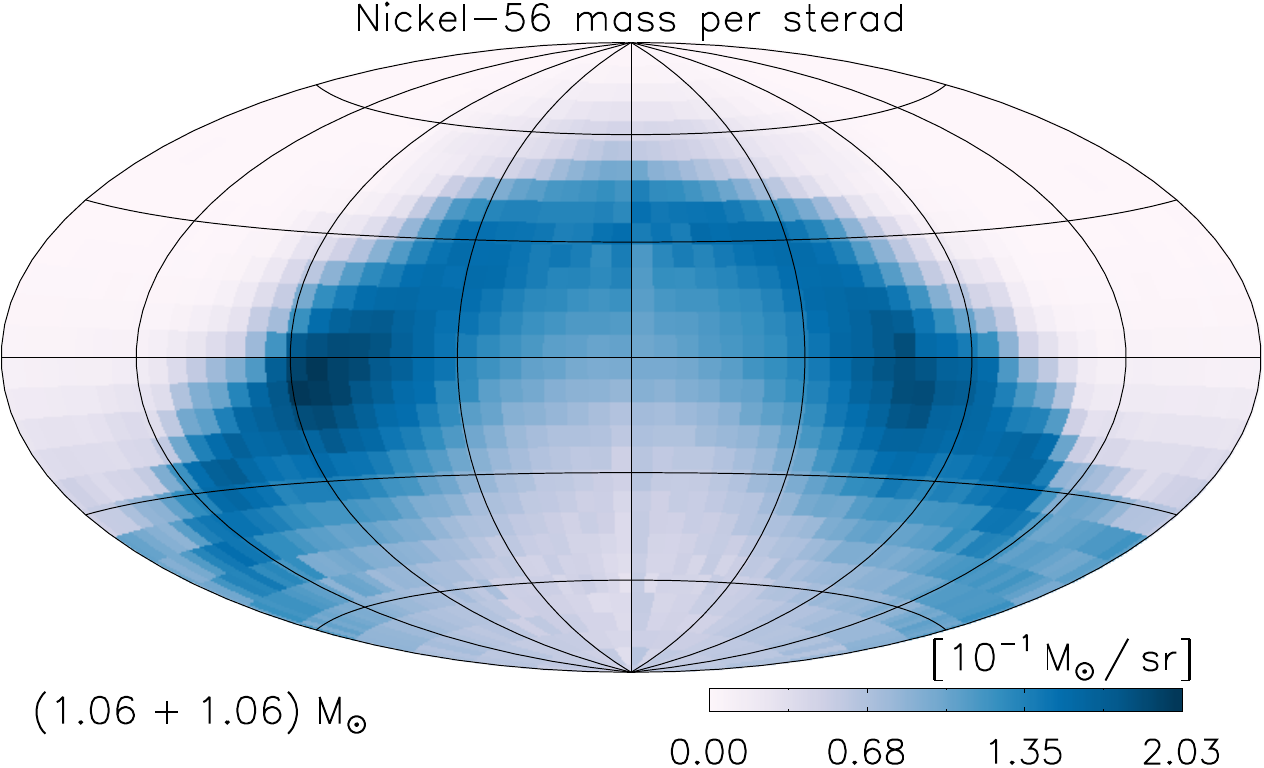} \\
\includegraphics[scale=\myscale]{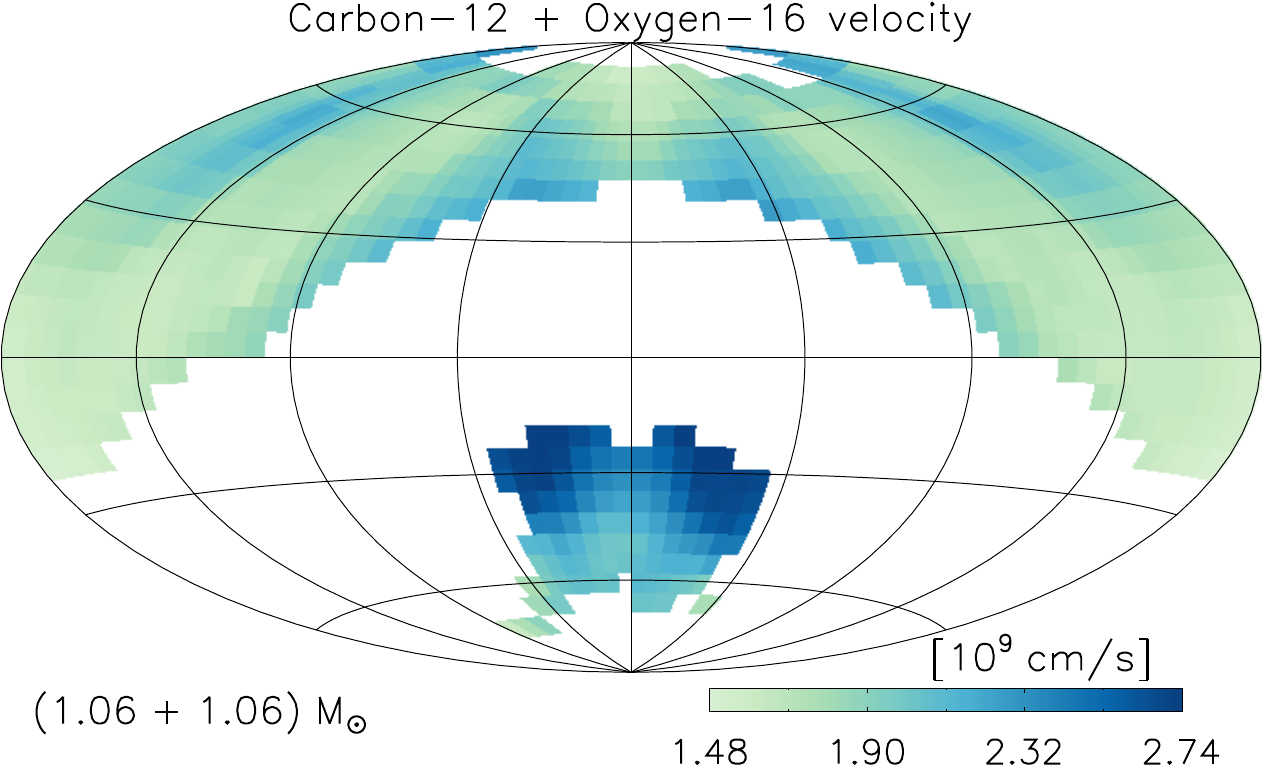} &
\includegraphics[scale=\myscale]{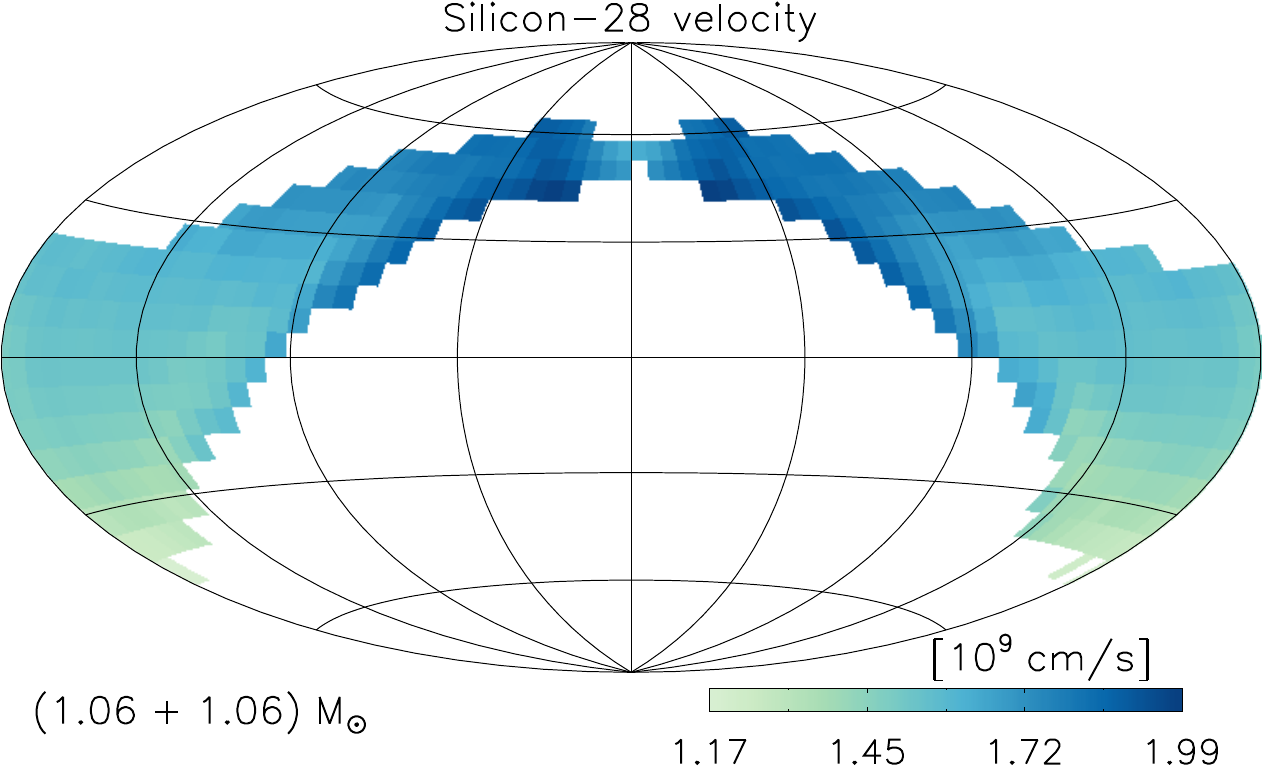} &
\includegraphics[scale=\myscale]{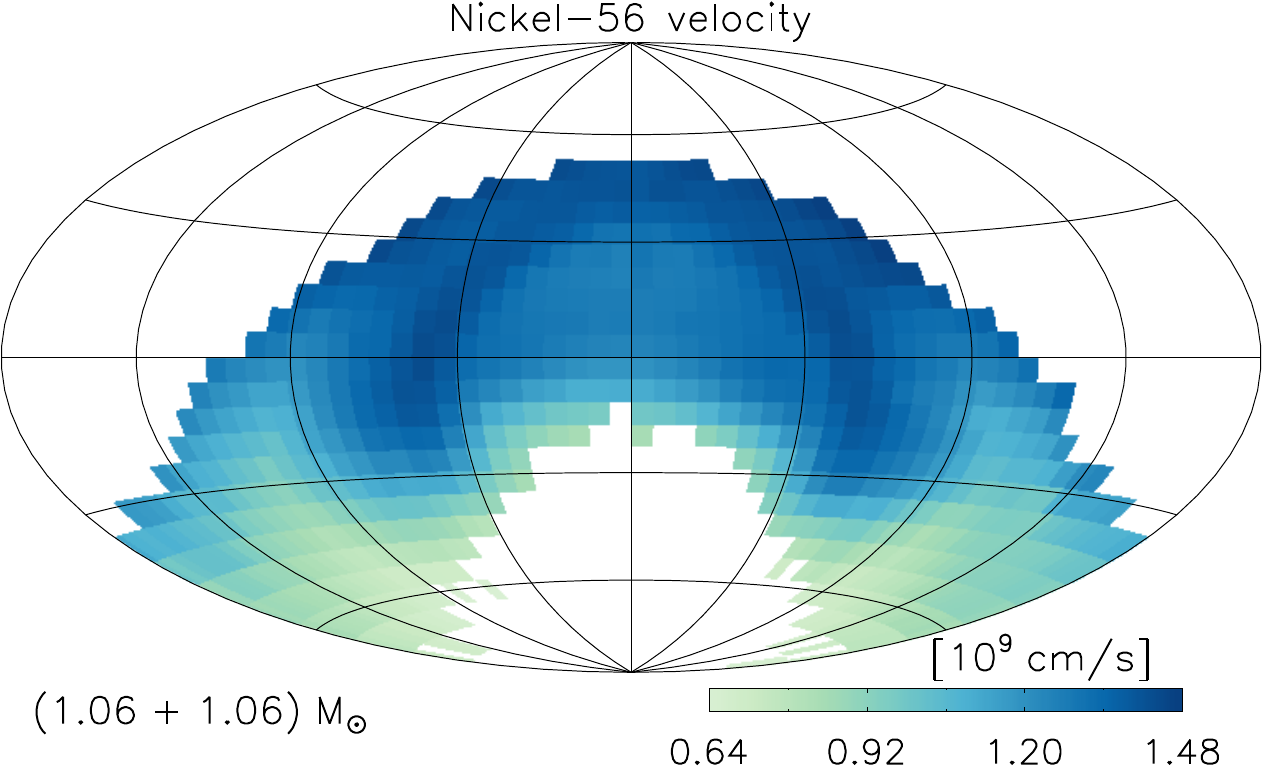} \\[\bigskipamount]
c) & & \\
\includegraphics[scale=\myscale]{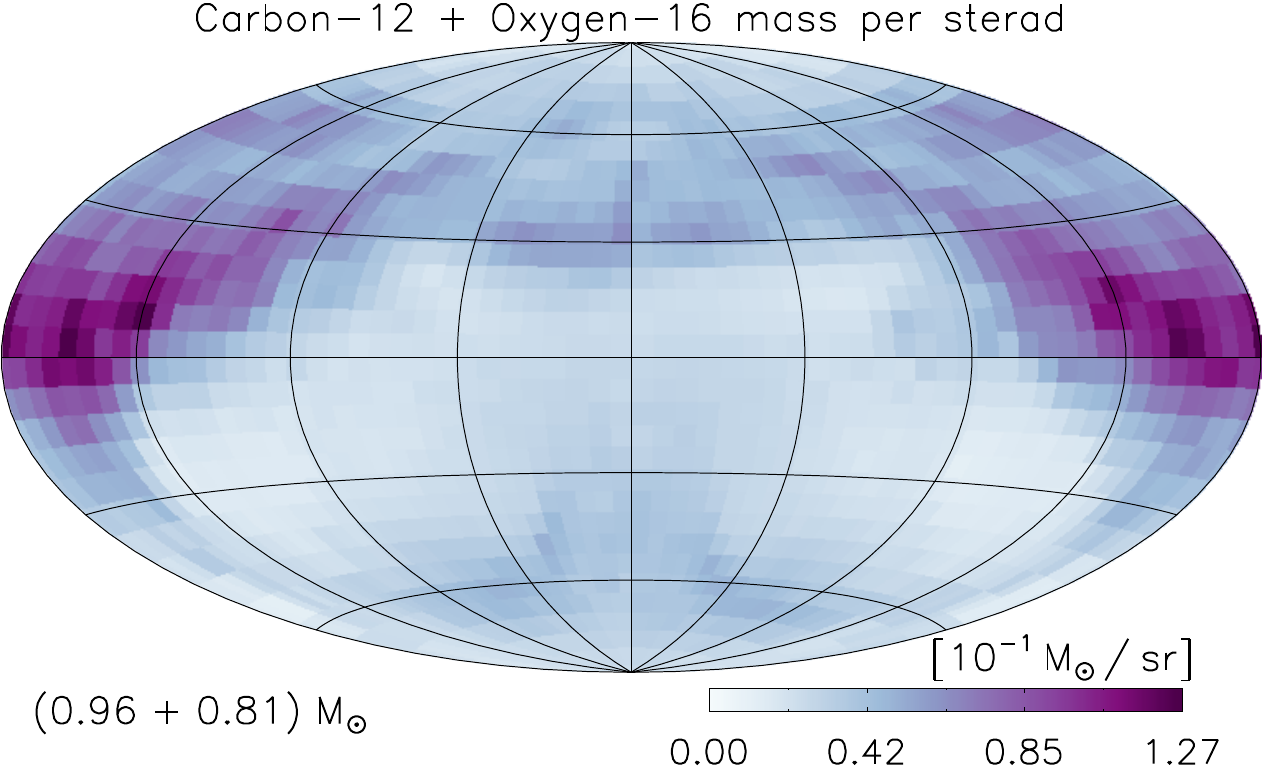} &
\includegraphics[scale=\myscale]{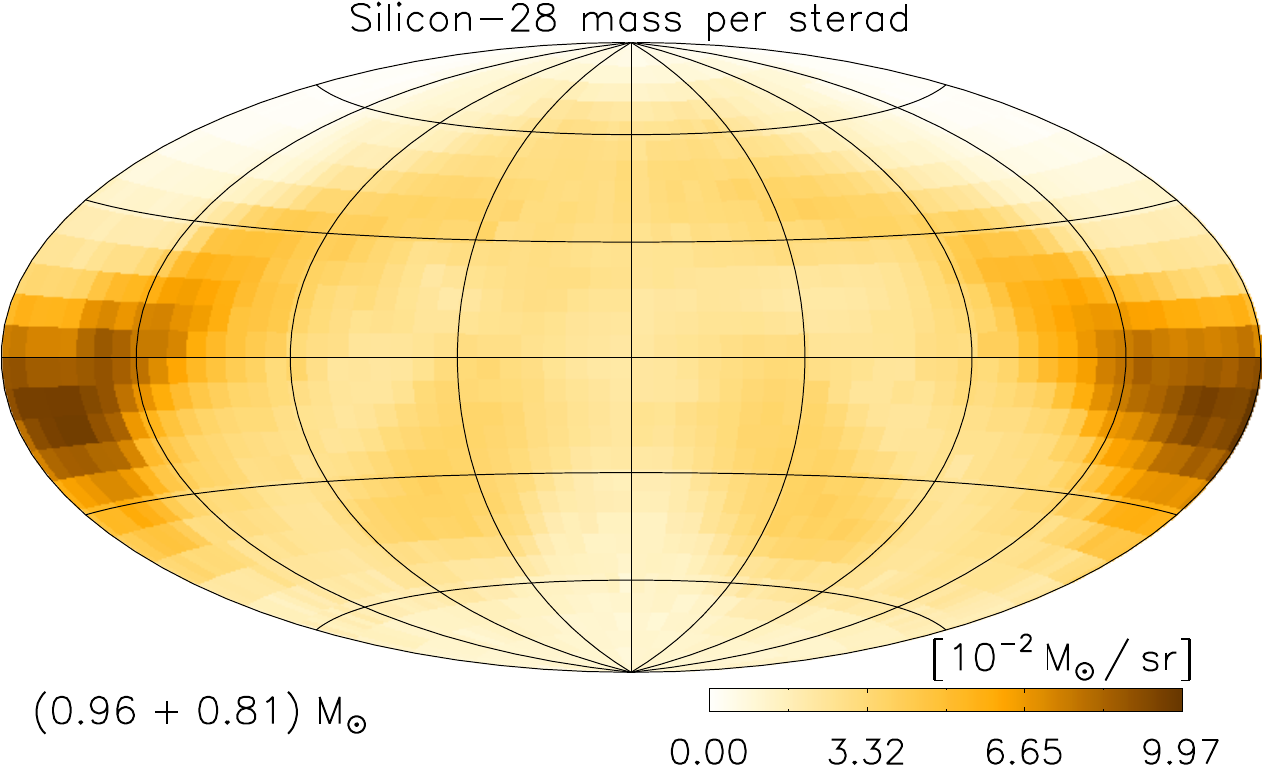} &
\includegraphics[scale=\myscale]{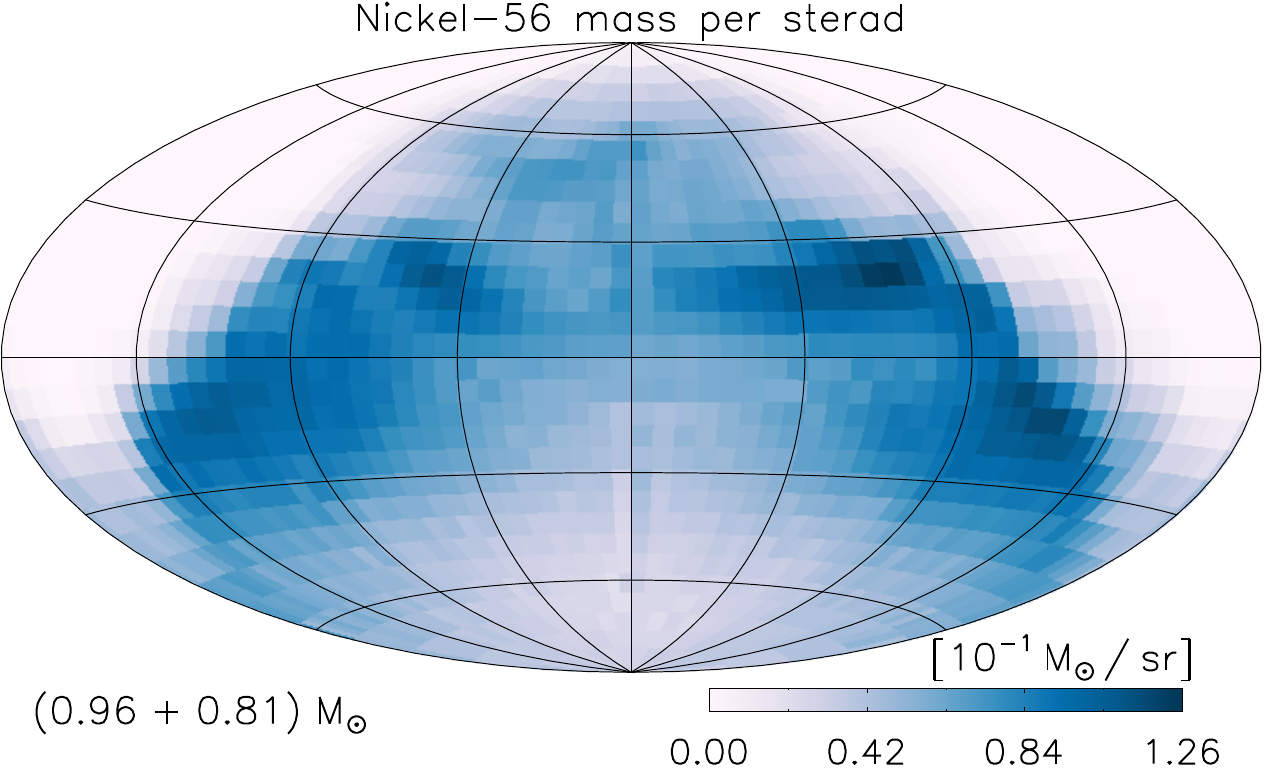} \\
\includegraphics[scale=\myscale]{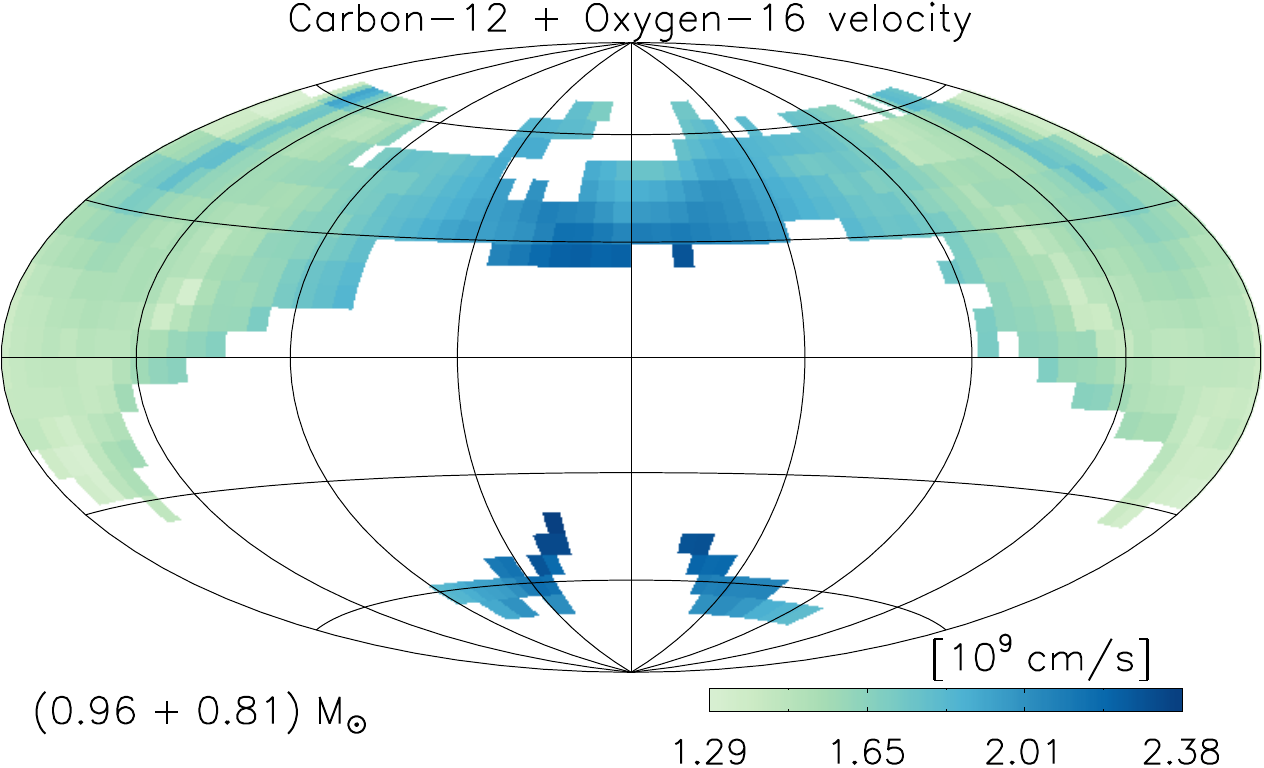} &
\includegraphics[scale=\myscale]{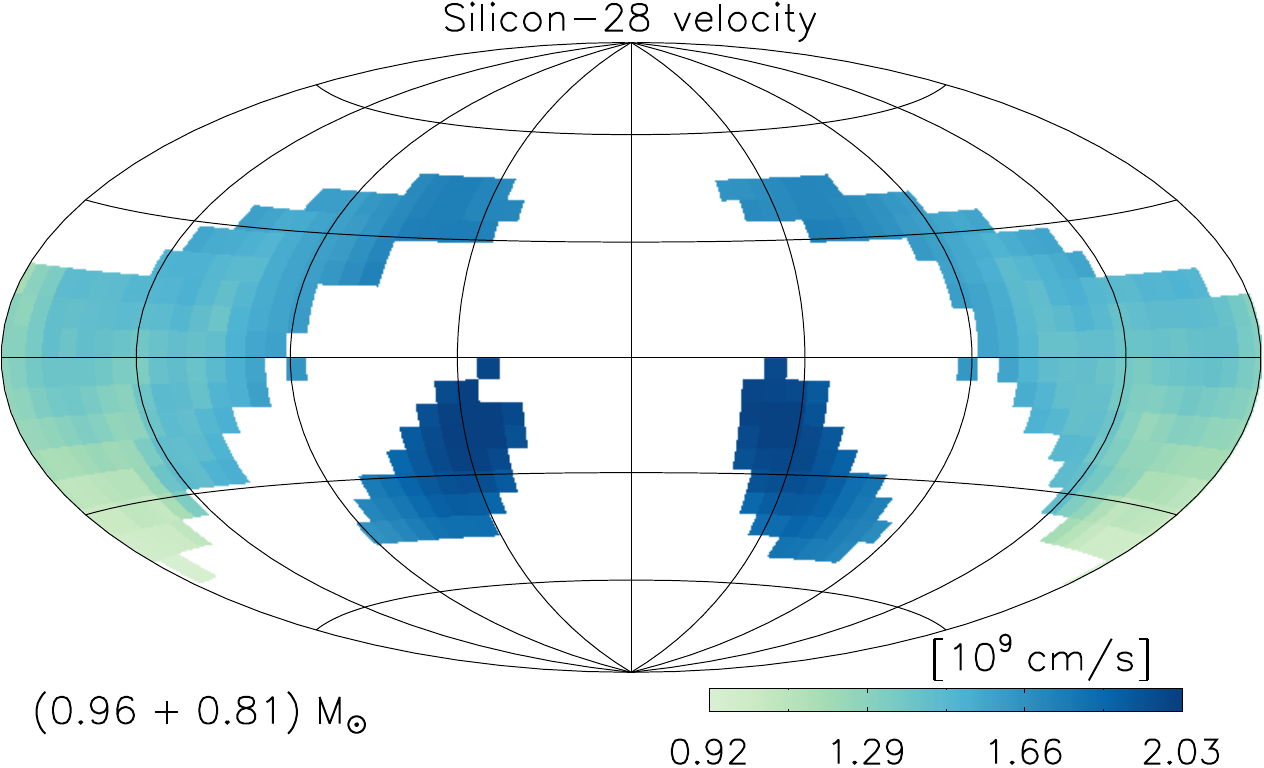} &
\includegraphics[scale=\myscale]{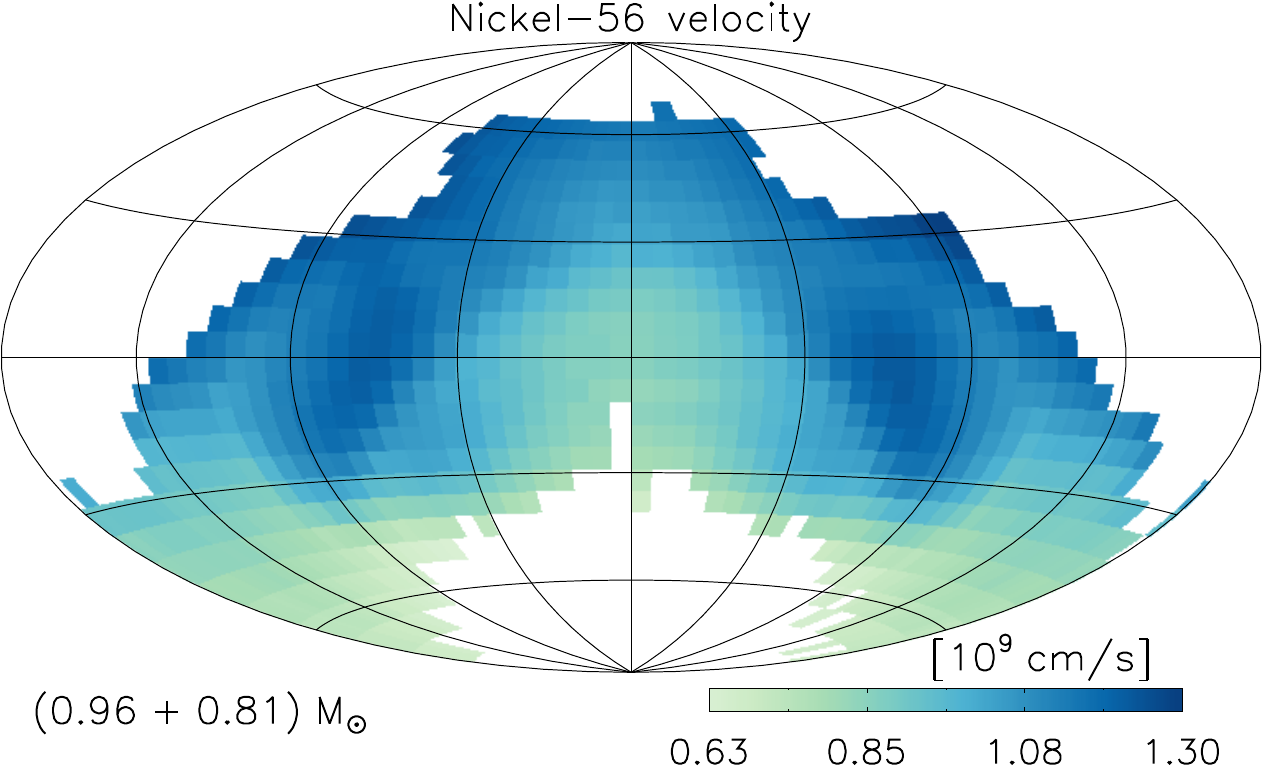}
\end{tabular}
\caption{\textit{Top row of each panel}: \el{C}{12}+\el{O}{16} (purple colors),
\el{Si}{28} (orange colors) and \el{Ni}{56} (blue colors) masses per unit of
solid angle as a function of spatial direction with respect to the explosion
center in a Hammer projection. See Figure~\ref{fig:projection} and
\S\ref{sec:sterad} for details on the orientation of the projection.
\textit{Bottom row of each panel}: Corresponding ejecta velocities, measured at
the smallest radius that contains 95\% of the mass in a given solid angle bin.
The velocities are only plotted for directions in which the mass density is at
least one third of the maximum (corresponding to the second label from the left
in the color bar of the respective mass density plot).}
\label{fig:sterad}
\end{figure*}

\subsection{Detonations}
\label{sec:detonations}

In all of the models, detonation waves emerge shortly
($\mathord{<}0.1\unit{s}$) after the simulations are resumed in CASTRO at the
state depicted in Figure~\ref{fig:initrho}.  The primary star is consumed by
the detonation on a time scale of half a second, see
Figure~\ref{fig:expl_t0403} for an example. The system rotates only a few
degrees during that time.
While a shock wave moves through the secondary star, it does not produce
additional \el{Ni}{56}. However, some \el{Si}{28} is produced in association
with the secondary and/or the accretion stream it feeds\footnote{While it is
not possible to unambiguously separate the yields of the two stars without
tracers, the formation of a \el{Si}{28} ``lump'' in the direction of the
secondary is evident.}.

The shape of the ejecta is governed by the anisotropic density of the medium
surrounding the primary.  After the detonation has consumed the primary, the
shock in the orbital plane accelerates most efficiently in directions ahead of
the primary, and least efficiently along the line connecting the two stars.  As
illustrated in Figure~\ref{fig:expl_t0403}, this anisotropic acceleration
causes the ejecta to become elongated in shape in the orbital plane.  The 3D
visualization of Figure~\ref{fig:densvol_t0401} demonstrates how the
acceleration of the \el{Ni}{56} ejecta is inhibited in some directions, forcing
it to become aspherical.

At about three seconds after the start of the detonation, the temperature has
become too low for nuclear reactions everywhere. The shapes of the ejecta are
still visibly evolving after that, albeit only slightly.  We stopped the CASTRO
simulations after $48\unit{s}$, at which time the ejecta are expanding
homologously.

\subsection{Yields}

Table~\ref{tab:yields} lists the isotope yields of all models.  Based on the
sum over the masses of all species except for the fuel elements \el{C}{12} and
\el{O}{16}, we categorize models \mtwo and \mone as super-Chandrasekhar mass
detonations, and model \mnine as sub-Chandrasekhar. The sum of the masses of
intermediate-mass elements (IMEs, $28 < A < 40$) and iron-group elements ($A
\geq 44$) are listed in a separate row for convenience.  The table also lists
the total energies (kinetic + internal + gravitational) at the beginning (when
the detonation sets off) and end of the simulations, as well as the energy
yields.

Figure~\ref{fig:massvel} indicates the location of these yields in the
homologously expanding ejecta by plotting the total masses and mass fractions
of different elements versus velocity.  The mass-weighted mean velocity of the
total ejecta is in the range of $(8.8\text{--}9.6) \times 10^8\cms$, with the
model \mone being the fastest, and model \mnine being the slowest.  The
mass-weighted mean velocity of the \el{Ni}{56} ejecta is smaller than that of
the IMEs in all cases.  Model \mnine stands out in having a high abundance of
IMEs without \el{Ni}{56} at the center.

\subsection{Ejecta Morphology}
\label{sec:morph}

The 3D structures of the coasting \el{Si}{28} and \el{Ni}{56} ejecta are shown
in Figure~\ref{fig:ejecta}.  In all cases, there is symmetry across the orbital
plane. There is no axisymmetry around the line that connects the stars at the
beginning of the detonation.  The extent of the \el{Si}{28} and \el{Ni}{56}
ejecta is anisotropic, reflecting that the ashes are accelerated most
effectively in directions ahead of the primary (with respect to the global
direction of rotation of the coalescing stars).

The ejecta of models \mone and \mnine are morphologically very similar. The
concave insides of the \el{Ni}{56} ejecta contain pockets of \el{Si}{28} as a
result of burning in the secondary star.  The fast-moving \el{Si}{28}
surrounding the \el{Ni}{56} ejecta presumably originates from the low-density
outer parts of the primary star.  Model \mtwo lacks most of this \el{Si}{28}
``frame''.

The ejecta of unburnt fuel are shown in Figure~\ref{fig:ejecCO}.  In models
\mone and \mnine, a torus-shaped feature surrounds the connecting line on the
side of the defunct secondary, with the fastest velocities in directions ahead
of the secondary.  We surmise that this is material expelled from the
shock-heated secondary.  The ejecta in model \mtwo form a preacher bench
instead of a torus.

\subsection{Direction-Dependence of Masses and Velocities}
\label{sec:sterad}

A quantitative representation of the ejecta is shown in Figure~\ref{fig:sterad}.
The plots in the top row of each panel represent the solid angle mass densities
of \el{C}{12}+\el{O}{16}, \el{Si}{28} and \el{Ni}{56} with respect to the
coordinate origin.  (Note that the actual center of the explosion, as defined
by $\vec{v}=0$ in the grid's frame of reference, is virtually coincident with
the coordinate origin at this stage. This is inevitable, because the explosion
center does not move while the length scales increase.) The total masses listed
in Table~\ref{tab:yields} can be retrieved by integrating over $4\pi$. The
orientation of the projection is sketched in Figure~\ref{fig:projection}.  The
polar axis for each plot is chosen to be the line connecting the centers of the
two stars (cf. Figure~\ref{fig:initrho}), with the ``north pole'' being on the
side of the secondary.  The prime meridian (thick blue line in the sketch) lies
in the orbital plane ahead of the primary.  Positive ``longitudes'' correspond
to positive rotations according to the right-hand rule. The $z$-axis
corresponds to $\pm90\degree$ longitude at the equator.

In all cases, the largest deposits of \el{Si}{28} are seen in directions behind
the defunct primary star near the orbital plane. This \el{Si}{28}, which is moving
slowly, corresponds to the central pocket inside the concave part of the
\el{Ni}{56} ejecta shown in Figure~\ref{fig:ejecta}.  Fast \el{Si}{28} can be
seen in other directions, albeit with much lower solid angle mass densities.

Large concentrations of \el{Ni}{56} are seen over a wide range of solid angles,
including directions near the $z$-axis (normal to the orbital plane) and behind
the defunct secondary star.  The mass densities for the unburnt fuel,
\el{C}{12}+\el{O}{16}, are in all cases high in directions ahead of the defunct
secondary star near the orbital plane. These are the directions in which no
\el{Ni}{56} and little \el{Si}{28} are seen.

\subsection{Energetics}

The total energy of the ejecta is positive in all cases (cf.
Table~\ref{tab:yields}), implying that there is enough energy for the ejecta to
expand indefinitely.  However, the spatial
distribution of the energies could allow for a gravitational collapse in some
local region.  We investigated the possibility of gravitational fallback of an
inner region onto the center of the explosion by calculating the total energy
as a function of radius, using 100 radial bins. The gravitational binding
energy is calculated to be
\begin{equation}
    E_\text{grav}(r) = - \frac{1}{2} G \int_{V_r} \de V \int_V \de V' \, \frac{\rho(\vec{r})
                         \rho(\vec{r}')}{\abs{\vec{r}-\vec{r}'}},
\label{eq:gravbind}
\end{equation}
where the first integral includes only the volume within a sphere of radius
$r$, here denoted as $V_r$, and the second integral includes the whole
space\footnote{This may be overestimating the chance for a fallback. One could
argue that the material at large radii ceases to contribute to the total
potential once the expansion of the inner part slows down at the onset of a
collapse, and the inner and outer parts become increasingly separated.  The
second integral would then have to include only $V_r$ instead of $V$, and the
energy needed to defy gravity would be less.}.  We thus estimate that the
potential fallback mass in all cases is insignificant, involving less than the
innermost $10^{-3}\,\Msun$.

The above estimate does not exclude gravitational fallback towards other points
in the ejecta.  An obvious candidate for such a point is the maximum of the
inner integral in equation~\eqref{eq:gravbind}, which is located
$\simm10^{10}\unit{cm}$ from the center (where the expansion velocity is
$\simm2.5\times10^{8}\cms$).  We repeated the above calculations with $r$
calculated around this point, using the kinetic energies measured by an
observer who moves with this point.  The potential fallback masses are again
small, less than $0.02\,\Msun$ in all cases.

Note that in the above, we have ignored the energies from radioactive decay and
electron recombination. The liberation of these latent energies could further
decrease the chance for fallback.  Also note that while our energetic argument
suggests that considerable gravitational fallback is unlikely, only
hydrodynamic calculations can definitely predict the long-term evolution of the
ejecta.

\section{Light Curves and Spectra}

\begin{figure*}[t]
\centering
\begin{tabular}{lcr}
\includegraphics[width=.28\linewidth]{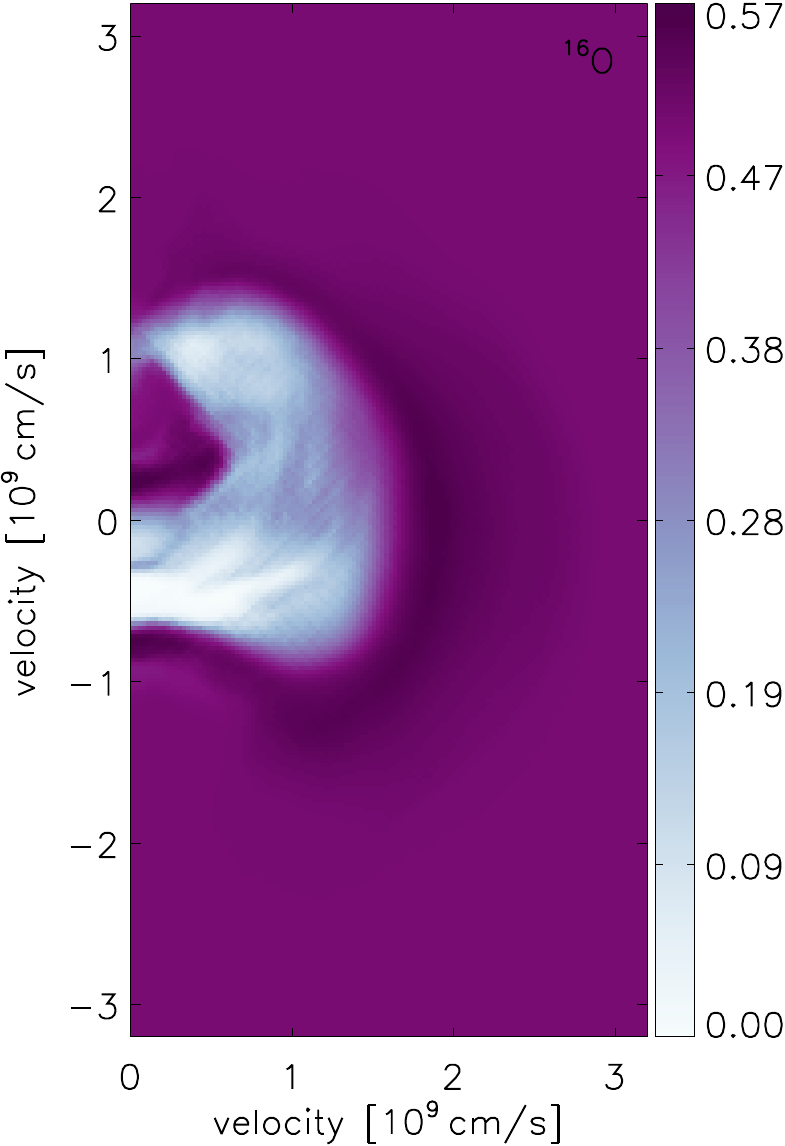} &
\includegraphics[width=.28\linewidth]{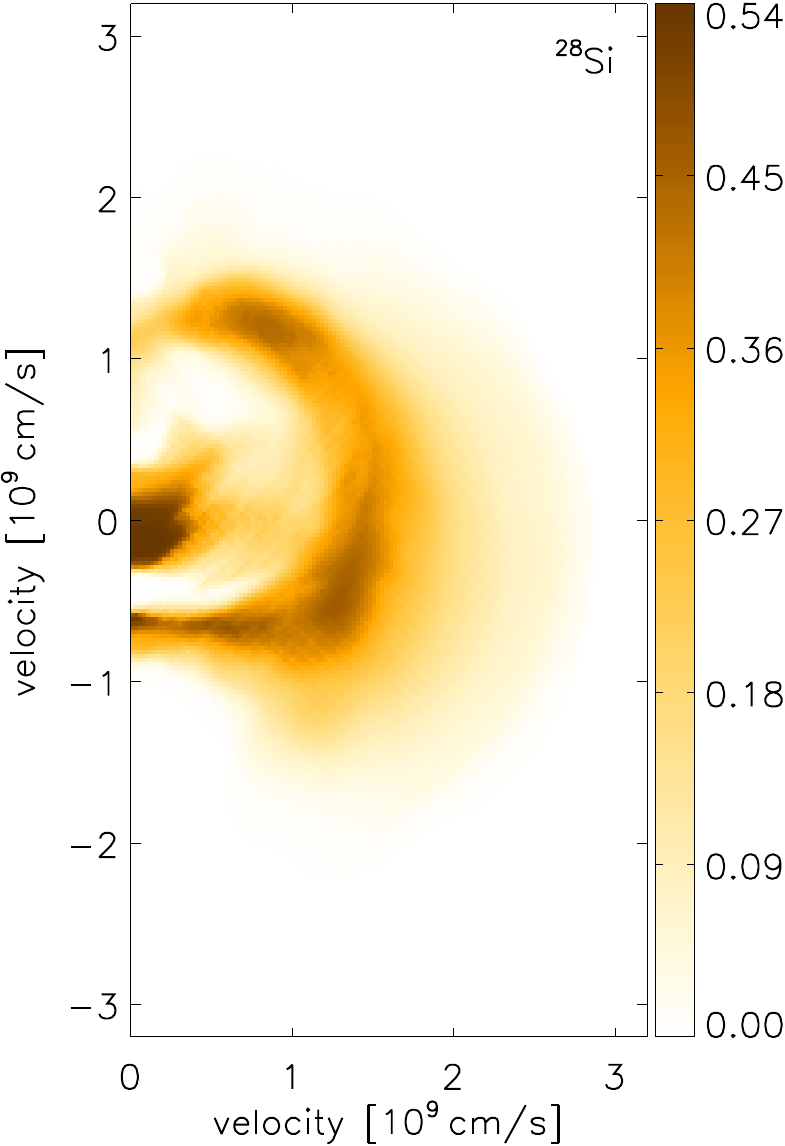} &
\includegraphics[width=.28\linewidth]{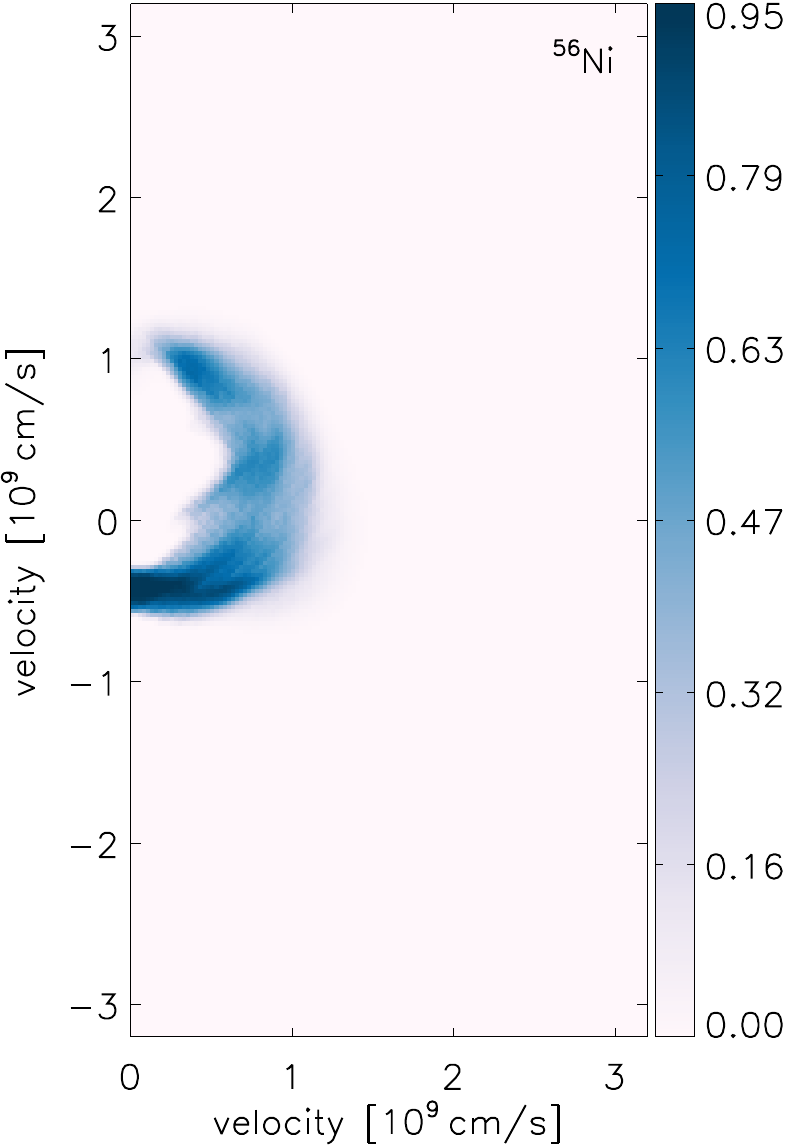}
\end{tabular}
\caption{Compositional structure in the azimuthally averaged version of model
\mnine.  The color coding shows the mass fractions of oxygen, silicon, and
radioactive nickel.}
\label{fig:2d_ejecta}
\end{figure*}

\begin{figure*}[t]
\centering
\includegraphics[width=.9\linewidth]{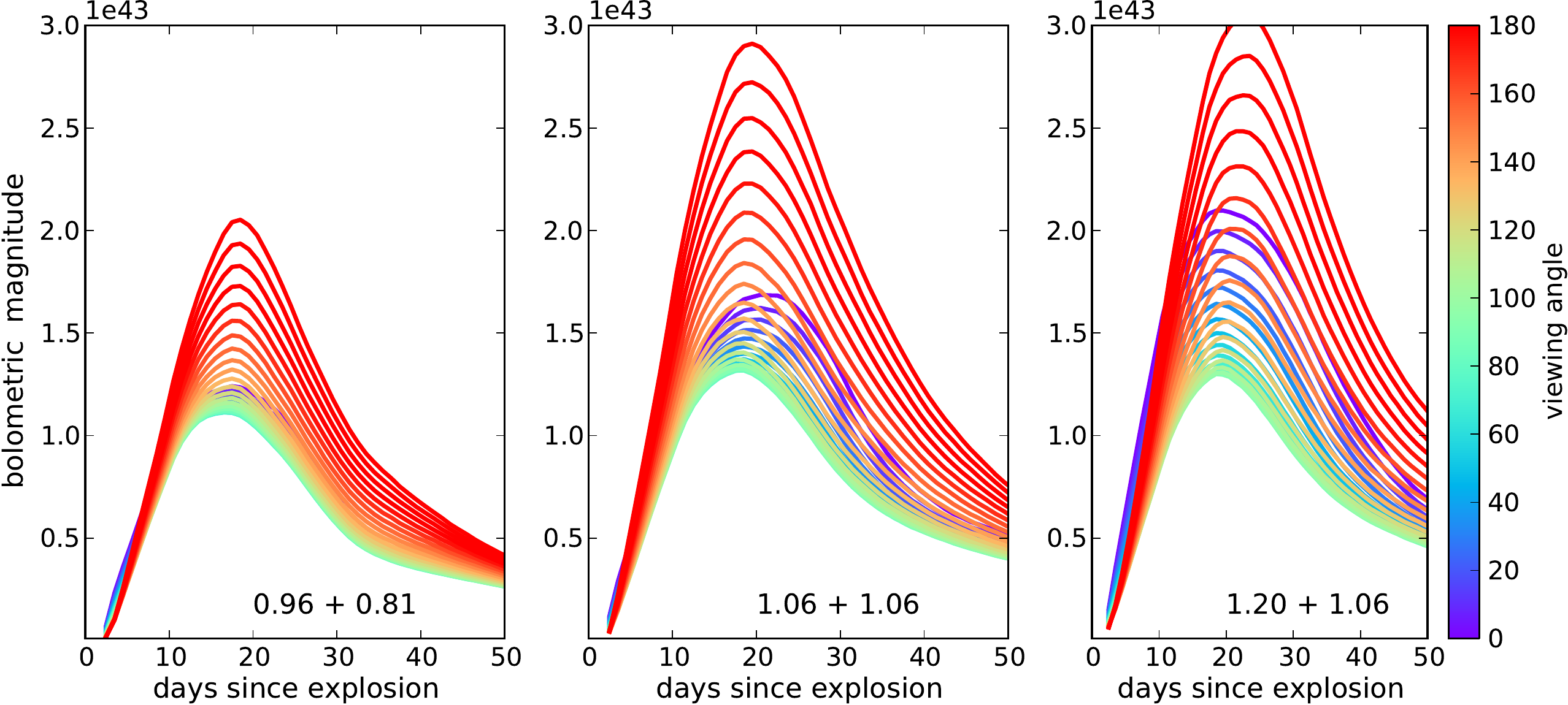}
\caption{Synthetic bolometric light curves of the merger models.  The color coding indicates the 
polar viewing angle.}
\label{fig:bol_lc}
\end{figure*}

\begin{figure*}[t]
\centering
\includegraphics[width=.9\linewidth]{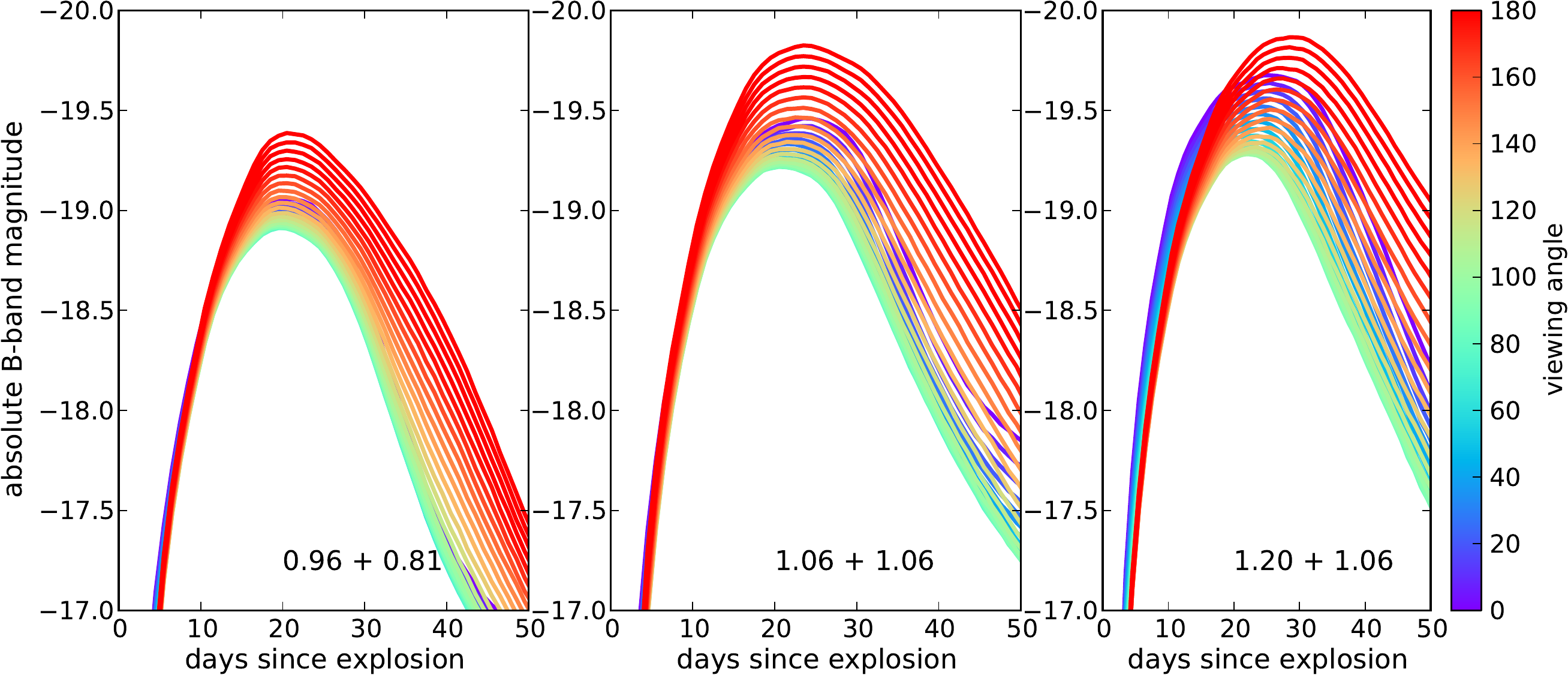}
\caption{Synthetic B-band light curves of the merger models.  The color coding indicates the polar
viewing angle.}
\label{fig:bband_lc}
\end{figure*}

\begin{figure}[t]
\includegraphics[width=\linewidth]{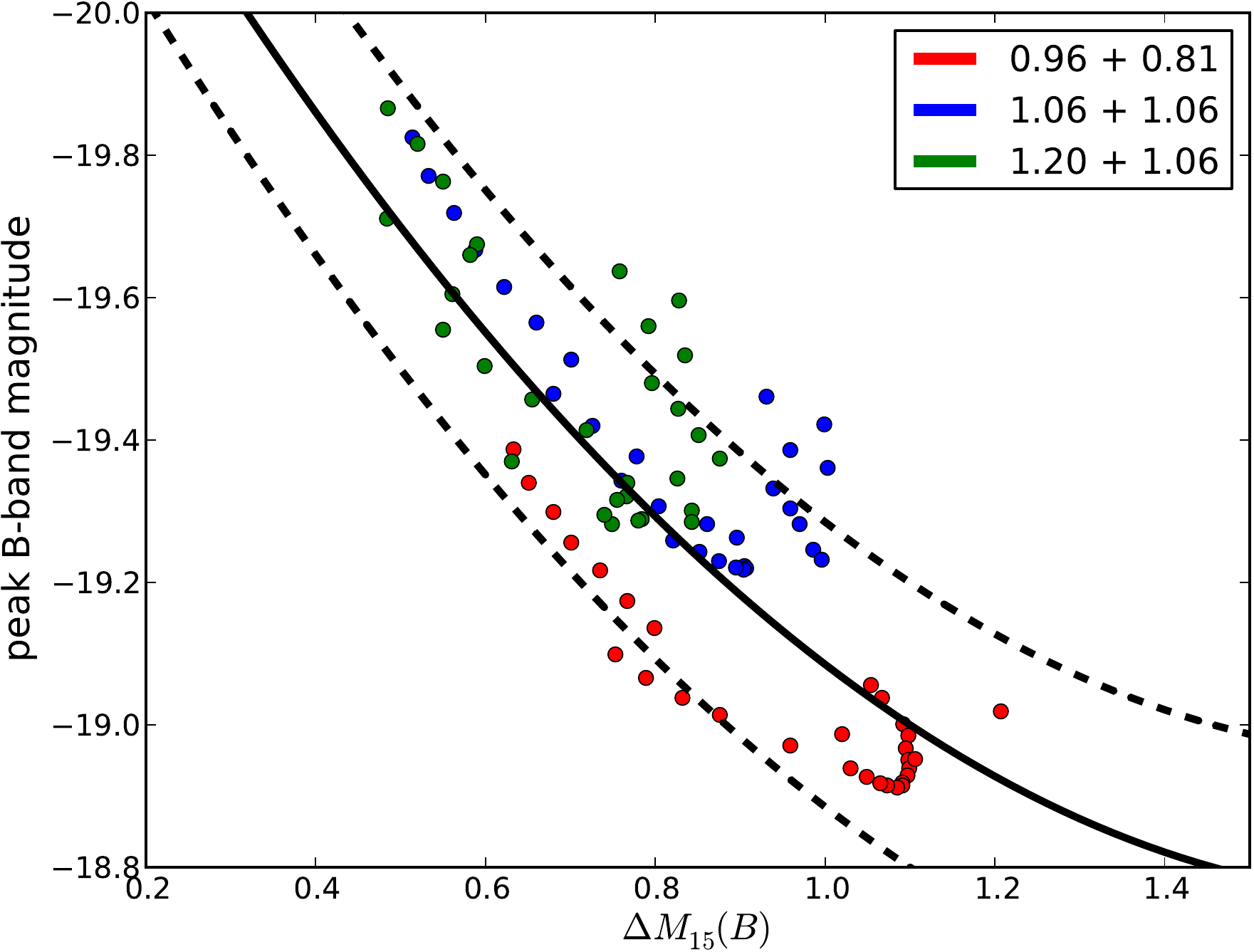}
\caption{Width-luminosity relation of the models.   The y-axis is the measured
peak B-band absolute magnitude and the x-axis is the  decline in magnitudes
from B-band  peak to 15 days later.  The individual points for each model
represent different polar viewing angles.  The black lines show the relation
from \cite{Phillips_1999}, $M_B = M_{B,0} + 0.786(\Delta M_{15}(B) - 1.1) -
0.633(\Delta M_{15}(B)  - 1.1)^2$.  The solid line uses $M_{B,0} = -19.0$,
while the upper and lower dashed lines use $M_{B,0} = -19.2$ and $M_{B,0} =
-18.8$, respectively.}
\label{fig:wlr}
\end{figure}

We generated synthetic light curves and spectra of our merger models by
post-processing the final output of the CASTRO hydrodynamical models with the
radiative transfer code SEDONA \citep{Kasen_MC}.  The parameters of the
transport calculations were equivalent to those of \citet{Kasen_2009} and adopt
the assumption of local thermodynamic equilibrium.    While not strictly
axially symmetric, the merger simulations do display a dominate axis of
symmetry.  The main dependence of the observables will therefore be on the
polar viewing angle as measured from the main symmetry axis.  
For ease in analysis and computation, we azimuthally averaged the
models into a 2D cylindrical coordinate system.  This was done by integrating
masses and momenta to obtain a grid with 128 radial and 256 vertical zones,
with the assumed axis of symmetry defined as the line connecting the centers of
the two stars when the detonation sets off.

Figure~\ref{fig:2d_ejecta} illustrates the ejecta composition structure
(plotted in velocity coordinates) for the azimuthally averaged model \mnine.
The geometry discussed in \S\ref{sec:morph} is clearly visible in this 2D
visualization, and will be useful for understanding the orientation dependencies
of the light curves and spectra.  The  \Nifs  has bowl-like shape, and is most
highly concentrated  in a thin pancaked region in the lower polar region.   The
velocity of the \Nifs in this lower polar region does not extend past $v
\approx 6{,}000 \kms$, whereas much higher \Nifs ejection velocities  $(v
\approx 10{,}000 \kms$) are seen the equatorial and upper polar regions.  The
maximum IME ejection velocities are also lower in the lower pole ($v \approx
10{,}000 \kms$) than the upper pole ($v \approx 15{,}000 \kms$).  The highest
IME velocities are in the equatorial regions, where they extend to $v \approx
20{,}000 \kms$.  A pocket of IME also exists near zero velocity, due to burning
of the secondary WD, and a pocket of low-velocity oxygen is visible as well.

\subsection{Synthetic Light Curves}

Figure~\ref{fig:bol_lc} shows the synthetic bolometric light curves, as a
function of the polar viewing angle, for each of the 3 merger models.  Due to
the strong ejecta asymmetry, the peak luminosity varies by a factor of nearly 2
or more with orientation.  For model \mnine, the  mean peak bolometric
luminosity (averaged over all polar angles) is $\Lbol = 1.32 \times 10^{43}
\ergss$, comparable to that of normal SNe~Ia.  However, for polar viewing
angles near $\theta = 180^\circ$ -- i.e., looking up along the z-axis in
Figure~\ref{fig:2d_ejecta} --  the  light curve is significantly over-luminous,
peaking at $2 \times 10^{43} \ergss$.   The anisotropy of the radiation is due
to the overabundance of \Nifs in the lower polar region, which results in a
greater energy deposition on one side of the ejecta, and a preferential escape
of photons in the downward direction.    The same orientation effects are
present in the more massive merger models, \mone and \mtwo, which have higher
total \Nifs masses and larger mean peak bolometric luminosities,  $\Lbol = 1.69
\times 10^{43} \ergss$ and $\Lbol = 1.77 \times 10^{43} \ergss$ respectively.
These two models are slightly over-luminous from most viewing angles, and from
the $\theta \approx 180^\circ$ they are extraordinarily bright, $\Lbol \approx
3 \times 10^{43} \ergss$, comparable  to the values seen in the rare class of
so-called ``super-Chandrasekhar'' mass events
\citep{Howell_2006,Hicken_2007,Scalzo2010,Taubenberger_2011,Silverman_2011}.  

Closer inspection of the synthetic light curves reveals that a large fraction
of the total luminosity is emitted in the ultraviolet bands ($\lambda \lesssim
3500\unit{\AA}$), and that the optical B-band light curves have somewhat more
moderate peak luminosities and orientation effects (Figure~\ref{fig:bband_lc}).
In model \mnine, the peak B-band magnitude varies by about $0.5\unit{mag}$ with
polar angle, from $M_B  = -19.3$ at $\theta = 180^\circ$ to $M_B \approx -18.8$
at $\theta \lesssim 90^\circ$.  This is within the range of normal SNe~Ia.
The peak B-band magnitudes of models \mone and \mtwo show a similar degree of
variation, reaching maximum values of $M_B \approx -19.8$ for $\theta =
180^\circ$.  These  brightest values are similar to, though a bit lower than
those  observed for the ``super-Chandrasekhar`` SNe~Ia, which are measured in
the range $M_B = -19.9$ to $-20.2\unit{mag}$ \citep{Taubenberger_2011}

Figure~\ref{fig:wlr} shows the  width luminosity relation (WLR) of the model
light curves, plotted as the correlation between the peak B-band absolute
magnitude and the B-band decline rate, $\Delta M_{15}(B)$.  The individual
points for a given model represent different viewing angles, and are seen to
generally follow the observed  ``broader equals brighter'' trend.  This
behavior is notable, given that in other asymmetric models  (in particular,
realizations of Chandrasekhar mass delayed-detonation explosions)  orientation
effects generally led to deviations from the WLR \citep{Kasen_Plewa_2007,
Kasen_2009, Sim_2013}.  The rough WLR trend obeyed by each model in
Figure~\ref{fig:wlr} can likely be explained by the  distribution of \Nifs
found in these merger simulations.  Because the velocity of  \Nifs is lowest in
the lower polar regions,  the line blanketing from iron group elements is
minimal when the SN is observed  near $\theta \approx 180^\circ$.    This
reduced degree of line blanketing at blue wavelengths results in a slower
B-band decline rate when observed from $\theta \approx 180^\circ$.   We note
that the bolometric light curves of the individual models  do not obey a strong
WLR (Figure~\ref{fig:bol_lc}), indicating that the B-band WLR relation of
Figure~\ref{fig:wlr} is indeed attributable to color evolution effects
\citep{Kasen_Woosley_2007}.

In detail, the model WLR relation  resembles the observed one of
\cite{Phillips_1999} with an assumed normalization of $M_B = -19.0$ at $\Delta
M_{15}(B) = 1.1$.  The scatter in the model relation is $\approx
0.2\unit{mag}$, which is comparable to, but slightly larger than that of the
observed sample without additional (e.g., color) corrections.  Given the
relatively high masses considered in our merger simulations, the models do not
populate the fast declining range of the plot  ($\Delta M_{15}(B) > 1.2$) and
are heavily weighted to  broader light curve events.  The brightest model
realizations have $M(B) = -19.8$ and decline very slowly, $\Delta M_{15}(B)
\approx 0.5\text{--}0.6$.  While these model points lie along the extrapolation
of the \cite{Phillips_1999} relation, the observed over-luminous
``super-Chandrasekhar'' SNe~Ia generally do not, having instead slightly faster
decline rates, $\Delta M_{15}(B) \approx 0.6\text{--}0.8$
\citep{Taubenberger_2011}.

\subsection{Synthetic Spectra}

\begin{figure}[t]
\includegraphics[width=\linewidth]{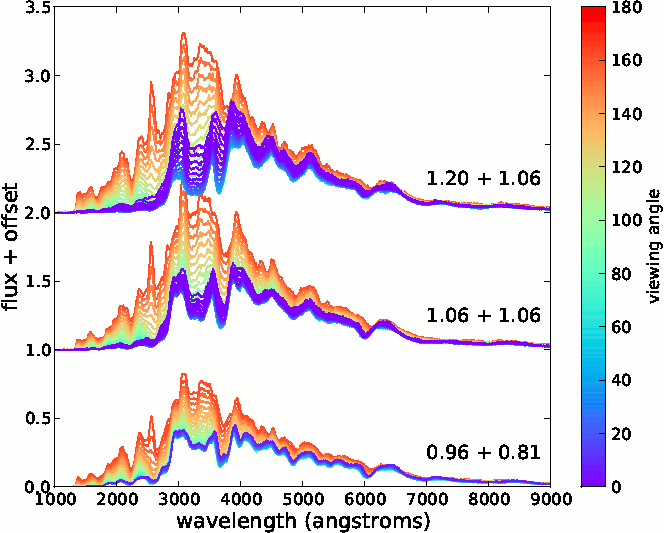}
\caption{Synthetic spectra at peak brightness for the models.  The spread for
each model shows the range of variation with viewing angle.}
\label{fig:spec_los}
\end{figure}

\begin{figure}[t]
\includegraphics[width=\linewidth]{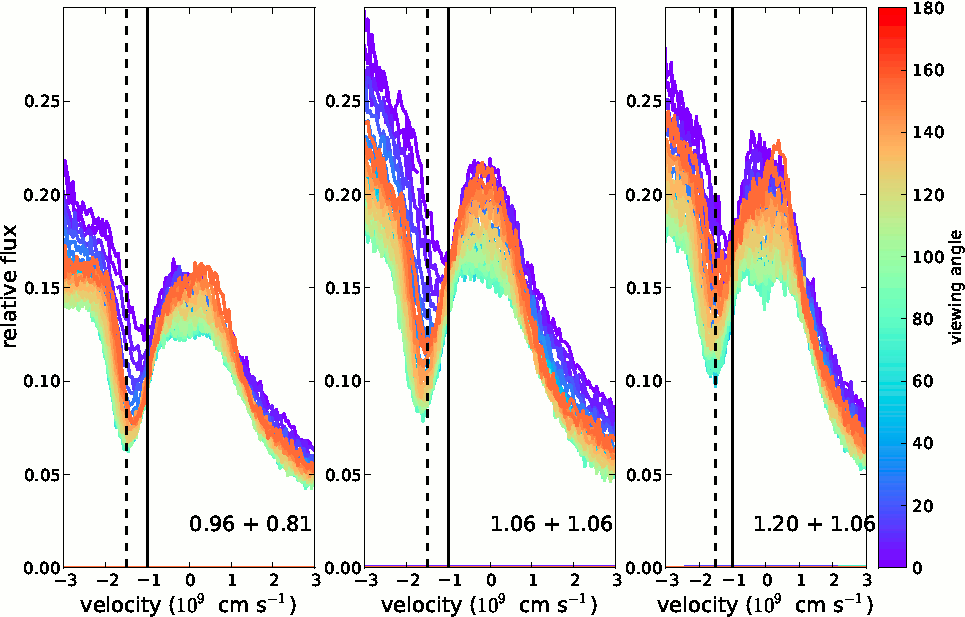}
\caption{Zoom in of the spectral region around the SiII absorption
feature, as seen from different viewing angles, for the models observed at peak
mean B-band brightness.  The x-axis shows the velocity relative to the SiII
gf-weighted rest wavelength of $6355\unit{\AA}$. The solid and dashed lines,
mark velocities of $10,000\kms$ and $15,000\kms$, respectively.  A CII
absorption feature is visible near zero SiII velocity in models \mnine and
\mtwo. }
\label{fig:sol}
\end{figure}

\begin{figure}[t]
\centering
\includegraphics[width=.9\linewidth]{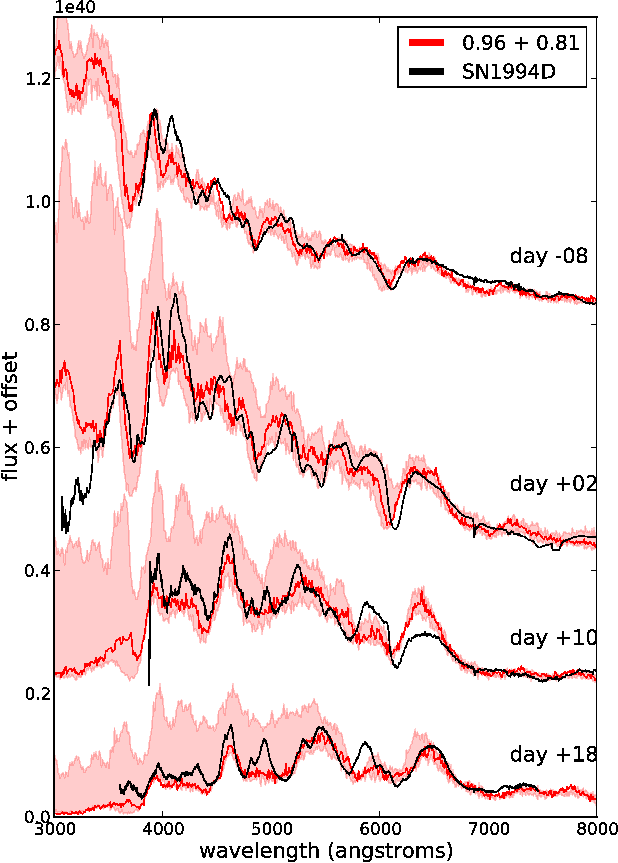}
\caption{Synthetic spectral time series for model 0.96 + 0.81, compared to
observations of the normal Type~Ia SN~1994D.  The shaded red region shows the
range of variation with viewing angle, and the solid red line marks the
equatorial $\theta = 90^o$ viewing angle.  We have normalized the model
spectrum at each epoch independently to provide an optimal comparison to the
observed spectral features.}
\label{fig:sn94d}
\end{figure}

Figure~\ref{fig:spec_los} shows synthetic spectra, as observed at peak
brightness, of each of the merger models.  The strong viewing angle
dependencies are apparent, and are most dramatic at bluer wavelengths.  For
model \mnine, the variation in the B-band flux ($\lambda \approx 4500\unit{\AA}$) is
a factor of $\approx 1.5$, while in the U-band ($\lambda \approx 3500\unit{\AA}$) the
variation is more than a factor of 2.  In the ultraviolet (UV) region of the
spectrum the flux varies with viewing angle by nearly a factor of 10.
Qualitatively similar orientation effects are seen in the synthetic spectra of
models \mone and \mtwo.  A similar variation in the B- and U-band flux with
viewing angle can be seen in the merger simulations presented by
\cite{2012Roepke}. 

The strong orientation effects of the model spectra can be understood, to first
order, by line blanketing effects.  The emission at blue wavelengths is
strongly influenced by the opacity of millions of blended, Doppler shifted iron
group lines, which are more highly concentrated at blue and ultraviolet
wavelengths.  In the lower polar regions of the ejecta, the synthesized iron
group (namely \Nifs) is restricted to a narrow range of velocities near $v
\lesssim 6000\kms$, which is  low compared to standard SN~Ia models like W7
\citep{Nomoto_1984}.  When viewed from polar angles near $\theta = 180^\circ$, the
line blanketing by iron group elements is therefore reduced, and the emergent
spectrum is significantly bluer.  From other angles, the \Nifs is distributed
across a greater range of velocities, and so provides higher line blanketing
and redder colors.  A similar effect of an asymmetric distribution \Nifs  on
SN~Ia colors was discussed in \cite{Foley_Kasen_2012}.

The geometrical distribution of the IMEs also has an influence on the spectral
line features seen at maximum light. Because the IMEs  have lower ejecta
velocities in the lower polar region, the absorption features of silicon,
sulfur and calcium lines all show lower Doppler shifts when observed from
$\theta \approx 180^\circ$.  Figure~\ref{fig:sol} shows a zoom in of the
hallmark Si~II feature (gf-weighted rest wavelength $6355$~\AA).  At peak
brightness, the velocity as measured from the minimum of the SiII absorption
feature varies by $\sim 40\%$, from $10{,}000\kms$ from $\theta = 180^\circ$ to
$14{,}000\kms$ from other angles.   

Clear carbon absorption features from CII  are also visible near the emission
peak of the 6355~\AA\ SiII feature in the maximum light spectra of models
\mnine and \mtwo  (Figure~\ref{fig:sol}).  This is a result of the relatively
large unburned carbon masses ($\simm 0.1\,\Msun$) left over from the secondary
star,  In the equal mass ratio merger \mone, less carbon remained unburned
($\simm 5 \times 10^{-2}\,\Msun$) and the CII absorption is less obvious at
these epochs.   The strength of the CII absorption varies with viewing angle,
given the asymmetric distribution of carbon in the ejecta.  Carbon lines have
been detected in normal SNe~Ia \citep{Branch_2003,Thomas_2011,Silverman_2012}.
In addition, ``super-Chandrasekhar" SNe~Ia frequently show significantly
stronger CII absorption.

On the whole, the synthetic spectra  of our merger simulations resemble those
of normal SNe~Ia.  Figure~\ref{fig:sn94d} compares the synthetic spectral time
series of model \mnine to observations of SN~1994D \citep{Hoflich_1995,
Meikle_1996,Patat_1996}.  Most of the major spectral features are reproduced,
although discrepancies can be noticed in the depth and velocity of individual
line profiles.  Similar  spectral fits were shown for the \mbox{1.1\,+\,0.9}
violent merger model of \citet{2011Pakmor} and \citet{2012Roepke}.  The overall
quality of the  fits is comparable to that found for multi-dimensional delayed
detonation explosions of Chandrasekhar mass WDs \citep{Kasen_2009,
2012Roepke,Sim_2013}.  The dispersion with viewing angle, however, is more
extreme in the merger model spectra, especially at UV wavelengths.  This is not
surprising given the greater ejecta asymmetry.

\section{Summary and Discussion}
\label{sec:discussion}

The merging of WDs was simulated for three different pairings of
progenitor masses, and the disruption of these systems by
thermonuclear detonations set off during the merging process was
studied.  All of the models considered generated energetic explosions
with high yields of \el{Ni}{56} and substantial amounts of IMEs, in
particular \el{Si}{28} and \el{S}{32}. The secondary star, which had
shed mass and was stretched by tidal forces at the time of ignition,
contributed to the nucleosynthesis of IMEs. In addition, it ejected
significant amounts of unburnt fuel.

The morphology of the ejecta was quite complex in general. The
presence of the companion star naturally breaks the spherical
symmetry, presumably more so than the non-central ignition.  In the
rotating system of coalescing WDs, the density drops more rapidly
ahead of the primary star, making the acceleration of the ashes more
efficient in this direction. This breaks the axisymmetry about the
line connecting the two stars.  Even though the rotational speed is
very slow compared to the detonation, it leaves an imprint on the
ejecta.  The high energies in the ejecta suggest that they will not
be subject to long-term gravitational interaction and that their
asymmetry will be retained indefinitely.

The total yields found in our calculations confirm that the violent
mergers of WDs can produce the nucleosynthesis needed for SNe~Ia.
However, not all kinds of pairings of WDs produce SNe~Ia such like
those that are observed. This raises the question as to why nature
would favor certain systems over others. We expect mergers involving
high-mass WDs to be rare for the simple fact that high-mass WDs are
rare.    Our \mtwo model
with carbon-oxygen composition clearly falls into this category.  
However, it has been argued that the WD primary may be able to 
grow via accretion during the binary evolution  \citealt{2013Ruiter}.
On the other hand, mergers of low-mass WDs probably do not
produce the necessary thermonuclear runaway that leads to a
detonation.  The ignition of the $(0.96+0.81)\Msun$ merger in this
work is marginal and uncertain.
We had to artificially induce a
detonation to make it work.  We also considered a $(1.06+0.64)\Msun$
merger, but that model did not reach the necessary detonation
condition (this model is instead presented in \citealt{Raskin_2013},
with a detonation being manually triggered in the post-merger state).
It seems fair to assume that models with lighter primaries in
combination with a $\simm0.64\,\Msun$ secondary will also fail to
ignite a prompt explosion.  Detonations in merging WDs with masses
lighter than in our \mnine model may therefore be unrealistic, unless
the ignition is aided by additional factors such as the presence of
helium \citep{2013Pakmor}.
Our findings are in line with a recent study by \citet{2013Dan} that
suggests a minimum total mass of $2.1\,\Msun$ for detonation in carbon-oxygen
merger systems, a condition satisfied only by our two heaviest models.

Another possible reason for diversity besides progenitor masses could be a
spread in the time when the detonation conditions are reached.  It is not clear
whether the evolutionary state at the time of the detonation is robustly always
the same for a similar pair of WD masses.  For instance, the $1.06\,\Msun$
secondary in combination with a $1.20\,\Msun$ primary has shed more mass than in
combination with a $1.06\,\Msun$ primary when the detonation sets off. Would the
state of the $1.06\,\Msun$ secondary in combination with a $1.12\,\Msun$ primary
lie in-between these two models? Our simulations suggest that these details
could have an effect on the production of IMEs as well as on the ejecta
morphology.  Many more models would have to be calculated to cover the possible
parameter space.

The synthetic light curves and spectra of our  \mnine merger model generally
resemble those of normal SNe~Ia.  Our results are  similar to those of
\cite{2012Pakmor}, who also found a favorable comparison to observation in
a \mbox{1.1\,+\,0.9} merger model.  Despite the smaller primary mass in our
model, the total amount of \Nifs synthesized ($M_\text{Ni} = 0.58\,\Msun$) was
only slightly lower than that of \cite{2012Pakmor} ($M_\text{Ni} =
0.616\,\Msun$).  This is most likely due to the different tables used to
calculate the nucleosynthesis in a detonation (see Table~\ref{tab:tabletest}).
In general, our simulations predict  greater \Nifs yields, which if correct
would imply that slightly lower mass WDs are sufficient for producing normal
SNe~Ia.    The  B-band light curve of our \mnine model has a mean (i.e., averaged
over all viewing angles)
peak magnitude of $M_B = -19$, a rise time of $t_B = 20$ days, and a decline
rate of $\Delta M_B(15) = 1.0\unit{mag}$, values which are very similar to
those found in the \mbox{1.1\,+\,0.9} model of \cite{2012Pakmor}. 

We have also considered the observables of violent mergers consisting of more
massive progenitors, namely \mone and \mtwo.  These models produced more \Nifs
($0.86\,\Msun$ and $0.99\,\Msun$, respectively) and will yield
over-luminous SNe~Ia.   Such events may be interesting in the context of the
very-luminous, ``super-Chandrasekhar"  SNe~Ia.   Due to the asymmetry of the
ejecta, the brightness of these merger models is boosted when viewed from
certain viewing angles (namely, those for which the bulk of the \Nifs is closer
to the observer).  As a result, models with a primary WD near or just
$1.2\,\Msun$ can, for some orientations, reach comparable brightness to the
observed events, even though the \Nifs masses are lower than what has been
empirically inferred on the basis of 1D models.  \citep{Hillebrandt_2007} had
previously suggested, in a different context, that asymmetry of the ejecta may
reduce the \Nifs  required to explain the ``super-Chandrasekhar"  SNe~Ia.  From
the brightest orientations ($\theta \approx 0^\circ$) the \mtwo\ model also
shows a slow light curve decline, a notable CII absorption feature,  and
relatively lower SiII absorption line velocities, all qualitatively consistent
with what is seen in the observed events.  
   
These violent merger models show strong variations with viewing angles.  These
are primarily due to the aspherical distribution of \Nifs, which leads to both
variations of the brightness and the degree of line blanketing with viewing
angle.  The range of variation in the ultraviolet is most extreme, with order
of magnitude flux variations seen.   SNe~Ia do show the greatest degree
of diversity at ultraviolet wavelengths, however the observed dispersion in
peak UV magnitude is not as large as in the models \citep{Brown_2010,
Cooke_2011}, We have also constructed a WLR for the violent merger models and showed
that, despite the strong dispersion of observables with viewing angle, the
models roughly lie along the observed trend.

Although the synthetic light curves and spectra we have calculated here are
generally consistent with observations of SNe~Ia, there may be other
observables of the models that are in conflict.  In particular, the strong
global asymmetry may lead to high levels of polarization, which would be
inconsistent with the generally low continuum polarization found in
observations \citep[see][and references therein]{Wang_Wheeler_2008}.  In
addition, the unusual chemical distribution of the ejecta, with a pocket of
inner oxygen surrounded by an outer bowl of \Nifs, may not be consistent with
nebular spectral of SNe~Ia, which typically show only moderately asymmetric
iron group lines and  no nebular oxygen emission.   Additional calculations
will be needed to explore these features and compare to observations.

We have also considered a late (post-merger) detonation of model \mnine,
details of which are presented in \citet{Raskin_2013}, and found higher
\el{Ni}{56} and lower IME yields as in the early (peri-merger) detonation
presented here.  The biggest difference, however, lies in the morphology of the
ejecta. This morphology could well be the decisive factor to determine whether
the double degenerate scenario is viable, and whether the detonation happens
during or after the merging process.

\begin{acknowledgments}
This research has been supported by the DOE HEP Program under contract
DE-SC0010676; the National Science Foundation (AST 0909129 and AST 1109896) and
the NASA Theory Program (NNX09AK36G).  DK is supported by a Department of
Energy Office of Nuclear Physics Early Career Award (DE-SC0008067).  RM
acknowledges support by the Alexander von Humboldt Foundation through the
Feodor Lynen Research Fellowship program.  We thank John Bell and Ann Almgren
for their major roles in developing the CASTRO code.  This research used
resources of the National Energy Research Scientific Computing Center, which is
supported by the Office of Science of the U.S. Department of Energy under
Contract No. DE-AC02-05CH11231.  This research used resources of the Oak Ridge
Leadership Computing Facility at the Oak Ridge National Laboratory, which is
supported by the Office of Science of the U.S.  Department of Energy under
Contract No. DE-AC05-00OR22725.
\end{acknowledgments}

\bibliography{ref}

\end{document}